\documentclass[twocolumn,prd,nofootinbib,aps,prl,floats,floatfix,amsmath,amssymb,longbibliography,secnumarab
ic]{revtex4-1} %
\usepackage[final]{graphicx}
\usepackage{hyperref}
\usepackage{amsmath}
\usepackage{bbm}
\usepackage{amsfonts}
\usepackage{amssymb}
\usepackage{latexsym}
\usepackage{graphicx}
\usepackage[english]{babel}
\usepackage{multirow}
\usepackage{float}
\usepackage{url}
\usepackage{slashed}
\usepackage{xcolor} 
\usepackage[utf8]{inputenc}

\def\sfrac#1#2{{\textstyle{#1\over #2}}}
\newcommand{\be}{\begin{equation}}
\newcommand{\ee}{\end{equation}}
\newcommand{\ba}{\begin{array}}
\newcommand{\ea}{\end{array}}
\newcommand{\bea}{\begin{eqnarray}}
\newcommand{\eea}{\end{eqnarray}}
\newcommand{\sss}{\scriptscriptstyle}

\newcommand{\nn}{\nonumber}
\renewcommand{\L}{{\sss L}}
\newcommand{\R}{{\sss R}}

\begin{document}

\title{\huge $R({K^{(*)}})$ from dark matter exchange}

\author{James M.\ Cline$^{\ (1,2)}$}
 \email{jcline@physics.mcgill.ca}
\author{Jonathan M.\ Cornell$^{\ (2)}$}
 \email{cornellj@physics.mcgill.ca}
\affiliation{$^{(1)}$ Niels Bohr International Academy \& Discovery Center,
 Niels Bohr Institute, University of Copenhagen,
Blegdamsvej 17, DK-2100, Copenhagen, Denmark}
\affiliation{$^{(2)}$ McGill University, Department of Physics, 3600 University St.,
Montr\'eal, QC H3A2T8 Canada}

\begin{abstract}

Hints of lepton flavor violation have been observed by LHCb in the
rate of the decay $B\to K\mu^+\mu^-$ relative to that of $B\to K
e^+e^-$. This can be explained by new scalars and fermions which
couple to standard model particles and contribute to these processes
at loop level.  We explore a simple model of this kind, in which one
of the new fermions is a dark matter candidate, while the other is a
heavy vector-like quark and the scalar is an inert Higgs doublet. We
explore the constraints on this model from flavor observables, dark
matter direct detection, and LHC run II searches, and find that, while
currently viable, this scenario will be directly tested by future
experiments.

\end{abstract}
\maketitle

%
\begin{figure}[b]
\hspace{-0.4cm}
\centerline{
\includegraphics[width=0.75\hsize]{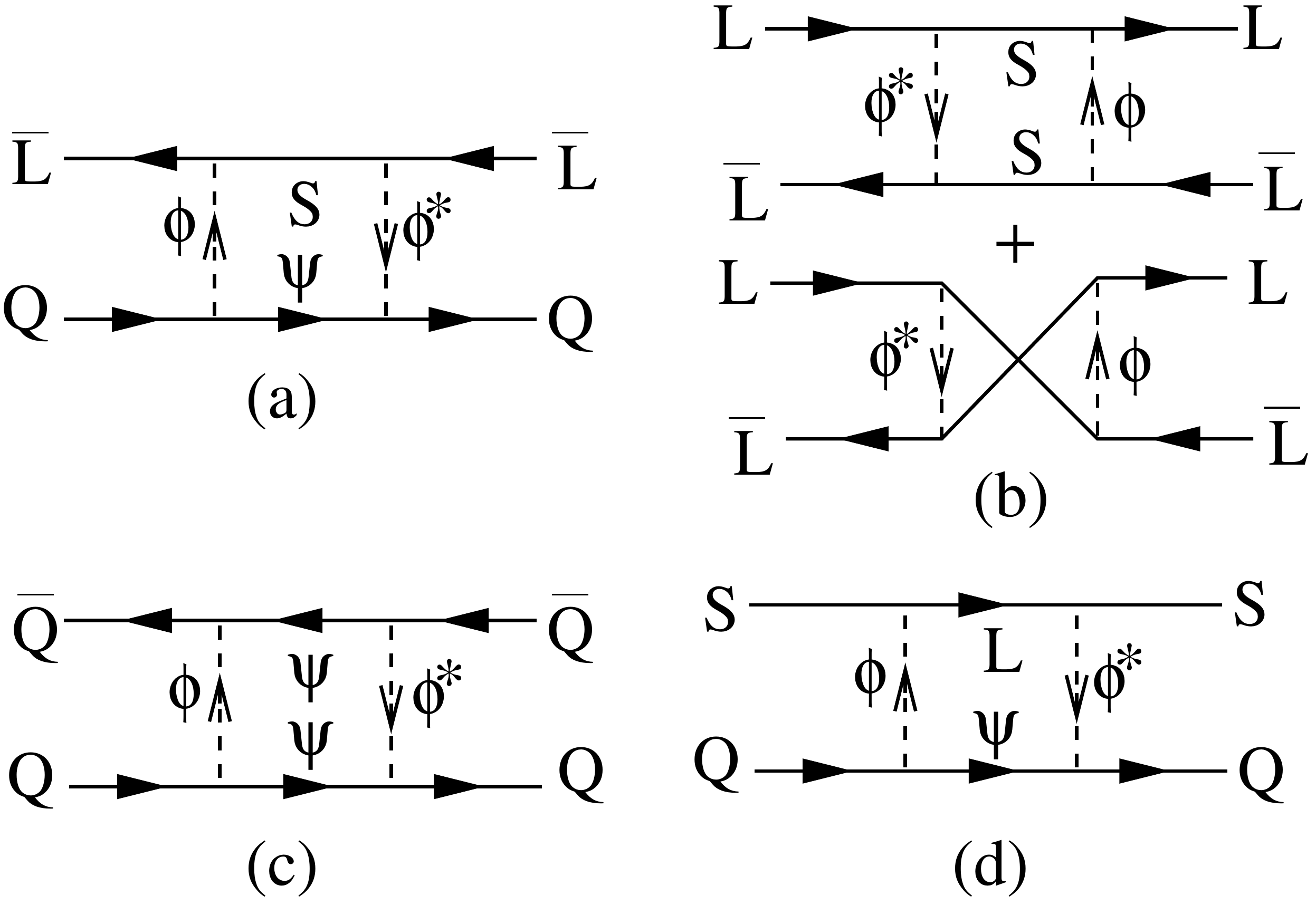}}
\caption{Diagrams leading to (a) $b\to s\mu\mu$, (b) $\tau\to 3\mu$, 
(c) $B_s$-$\bar B_s$ mixing and (d) dark matter scattering on 
quarks. Arrows on the $\phi$ scalars show the
flow of SU(2)$_\L$ quantum number, presumed to not be carried by
$S$ or $\Psi$.}
\label{box}
\end{figure}
%

{\bf Introduction.}  The LHCb experiment has observed intriguing 
deficits in $R(K)$ and $R(K^*)$, defined as the ratio of  branching
ratios $B(K^{(*)}\to \mu^+\mu^-)/B(K^{(*)}\to e^+e^-)$ 
\cite{Aaij:2014ora,Aaij:2017vbb}. 
These ``hadronically clean'' ratios are free from theoretical uncertainties
in hadronic matrix elements, which cancel out \cite{Hiller:2003js}.
In the
standard model (SM) it is expected that $R(K^{(*)})=1$
\cite{Bordone:2016gaq}, while
experimentally deficits of approximately 20\% are observed.   Although
the significance in either observation $K$ or $K^{(*)}$ is not high,
model-independent fits to both data, and possibly including quantities
more sensitive to hadronic physics, including  $B_s\to \mu^+\mu^-$, $B_s\to \phi\mu^+\mu^-$ and the
angular observable $P_5'$, 
indicate a higher significance of $\sim 4\sigma$
\cite{Capdevila:2017bsm,Geng:2017svp,Ciuchini:2017mik,Hiller:2017bzc,Altmannshofer:2017yso} 
Ref.\ \cite{DAmico:2017mtc} shows that the best fits
and significance do not change appreciably whether one includes the
hadronically sensitive observables or not, and
that it is possible to find a good fit to the data by including a
single dimension-6 operator in the effective Hamiltonian,
\be
	H_{\rm eff} \ni {\cal O}_{b_\L\mu_\L} = 
{1\over\Lambda^2}(\bar s_\L\gamma_\alpha b_\L)(\bar\mu_\L\gamma^\alpha \mu_\L)
\label{dim6op}
\ee
with $\Lambda \cong 31\,{\rm TeV}$, which is approximately
$-0.15$ times the SM contribution at one loop.

The new physics contribution (\ref{dim6op}) can be obtained from
tree-level exchange of a heavy $Z'$ vector boson 
\cite{DiChiara:2017cjq,Chiang:2017hlj,Dalchenko:2017shg,Agrawal:2017cbs,
Romao:2017qnu,Faisel:2017glo,Falkowski:2018dsl,Kohda:2018xbc} or leptoquark \cite{Chen:2017hir,Crivellin:2017zlb,Cai:2017wry,
Chauhan:2017ndd,Diaz:2017lit,Dorsner:2017ufx,Crivellin:2017dsk,Aloni:2017ixa,
Assad:2017iib,DiLuzio:2017vat,Calibbi:2017qbu,Dey:2017ede,Chauhan:2017uil,
Muller:2018nwq,Dorsner:2018ynv,Hiller:2018wbv,Fajfer:2018bfj,Monteux:2018ufc}, or
through loop effects of new particles.  In ref.\ 
\cite{Arnan:2016cpy}, an exhaustive classification and study of the
simplest loop models was carried out,  where it was shown that one
needs either two new scalars and one new fermion, or two new fermions
and one new scalar, to explain the $B$ decay anomalies.  Many possible
quantum numbers of the new particles are possible.  Here we note that
these include cases where one of  them can be neutral under the SM
gauge interactions, opening the possibility that it could be dark
matter (DM), and thus allowing the model to explain two observed phenomena 
requiring new physics.

We prefer to minimize the number of new scalars so there is just one,
thereby allowing the DM candidate to be one of the new fermions.\footnote{Ref.\ 
\cite{Gripaios:2015gra} focuses on the opposite choice, and observes that
the possible scalar dark matter candidate cannot satisfy direct
detection constraints because of its coupling to $Z$.
Previous
attempts to connect $R(K^{(*)})$ to  dark matter can be found in
refs.\ \cite{Sierra:2015fma,Belanger:2015nma,Celis:2016ayl,
Altmannshofer:2016jzy,Cline:2017lvv,Baek:2017sew,Cline:2017aed,
Sala:2017ihs}.  In addition,
refs.\ \cite{Kawamura:2017ecz,Chiang:2017zkh} recently studied models similar to ours, but in which
the DM is chosen to be a new scalar.  These studies do not fully consider the impact of the Higgs portal coupling $\lambda|H|^2|\phi|^2$ on the
DM relic density and direct detection.  In ref.\ \cite{Bhattacharya:2015xha} it was shown
that $\lambda$ tends to dominate over any other new physics
effects.  Even if it vanishes at tree level, the one-loop correction
tends to be too large to ignore without fine tuning.
}  Fermionic dark matter is free from relevant Higgs portal couplings,
making for a more predictive theory in which the dark matter
properties are
determined by the same couplings that explain the flavor anomaly. 
It will be shown that considerations of the dark matter relic density and direct
detection give interesting additional restrictions on the model,
and that it is also constrained by existing LHC searches as well as
flavor-changing neutral current processes.  The model therefore has
high potential for discovery by a variety of complementary
experimental searches.

\begin{table}[b]
\begin{tabular}{|c|c|c|c|c|c|c|c|}
\hline
 & SU(3) & SU(2)$_L$ & U(1)$_y$ & U(1)$_{\rm em}$& $Z_2$ & $L$ & $B$\\
\hline
$\Psi$ & $3$ & $1$ & $+2/3$ & $+2/3$ & $-1$ & $-1$ & $+1/3$\\
$S$    & $1$ & $1$ & $0$ & $0$ & $-1$ & $0$ & $0$ \\
$\phi$ & $1$ & $2$ & $-1/2$ & $(0,-1)$ & $-1$ & $+1$ & $0$\\
\hline
\end{tabular}
\caption{Quantum numbers of new physics particles, including
accidental $Z_2$ discrete symmetry that insures stability of the
dark matter $S$,  baryon ($B$) and lepton ($L$) number.  SM particles do not transform under the $Z_2$.}
\label{tab1}
\end{table}

{\bf Model and low-energy effective theory.}  We introduce a 
Majorana fermionic DM particle $S$, 
a vectorlike heavy quark $\Psi$ that carries SM color and hypercharge, and a
scalar $\phi$ that is an inert SU(2)$_L$ doublet.  The quantum numbers are
shown in table \ref{tab1}.  The only couplings of the new fields to
SM particles allowed by gauge and global symmetries (see table
\ref{tab1}) are 
\bea
	-{\cal L} &\ni& \tilde\lambda_i \bar Q_{i,a} \phi^a \Psi + 
	\lambda_i \bar S\phi_a^* L_i^a\nn + {\rm H.c.}\\
 &+& \lambda_{H,1}|H|^2|\phi|^2
	+ \lambda_{H,2} |H^\dagger\phi|^2 
\label{simple_mod}
\eea
where $Q,L$ are the SM quark and lepton doublets, $a$ is the SU(2)$_L$ index and $i$ is the flavor index. The
relevant interactions at low energy are generated at one loop and thus
require sizable couplings.  Since there is no flavor symmetry, we
will see that this model lives in a corner of parameter space where
meson mixing constraints are nearly saturated.   In a more complete
model, the global symmetries could be an accidental consequence of 
a spontaneously broken gauge symmetry under which the new physics
particles are charged. 

\begin{figure}[b]
\hspace{-0.4cm}
\centerline{
\includegraphics[width=0.45\hsize]{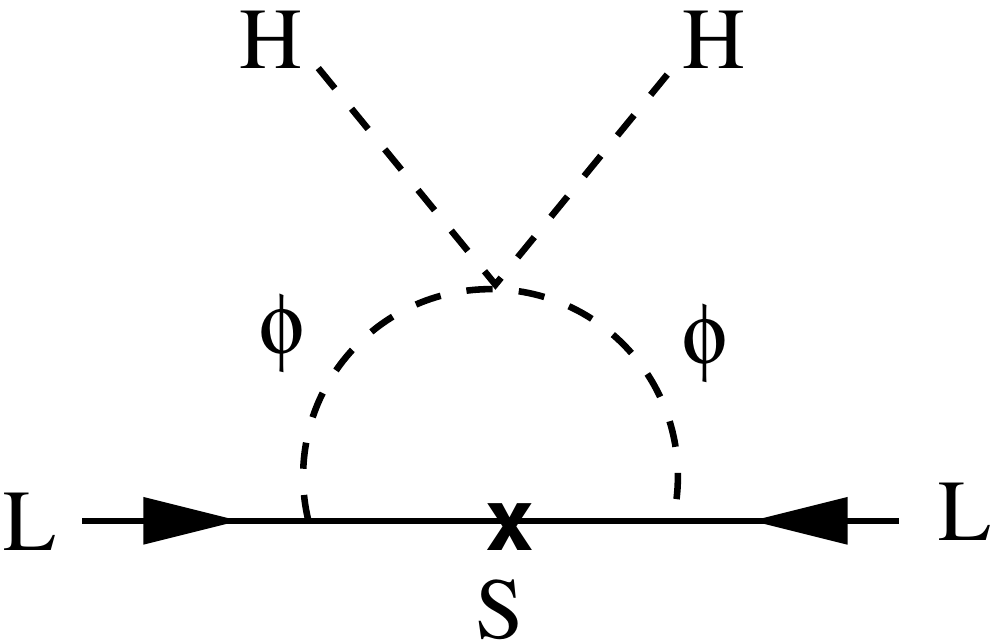}}
\caption{Loop-induced contribution to light neutrino Majorana mass.}
\label{loop}
\end{figure}

The Higgs portal couplings $\lambda_{H,i}$ play no important role in
the following;  $\lambda_{H,1}$ gives an overall shift to $m^2_\phi$
after electroweak symmetry breaking, while $\lambda_{H,2}$ splits the
charged and neutral components of $\phi$ by a small amount (relative
to $m_\phi^2$ as constrained by LHC searches).  A coupling of
the form 
\be
\lambda_{H,3}
(H^\dagger\phi)^2 +{\rm H.c.}
\ee
violates lepton number conservation, as can be seen from the
charge assignments in table \ref{tab1}.  (Notice that  $S$ cannot be
assigned lepton number since it is Majorana.)  
Of course one expects
that $L$ is only an approximate symmetry, if neutrinos have Majorana
masses, which constrains the size of $\lambda_{H,3}$.  In fact
this operator could be the origin of one of the neutrino masses through the 
loop diagram shown in fig.\ \ref{loop}, with mass matrix
$\delta m_{\nu,ij} \sim
\lambda_i\lambda_j\lambda_{H,3} m_S v^2/(16\pi^2 m_\phi^2)$
(where $v=246\,$GeV), which
has a single nonvanishing eigenvalue given by the trace.\footnote{A
more complicated model with two or more flavors of dark matter would
allow for nonsingular mass matrices.}  If $m_{\nu,3} = 0.05\,$eV for
example, $\lambda_{H,3}\sim 10^{-9}/\sum_i\lambda_i^2$.

%
%

To make definite predictions from (\ref{simple_mod}), we must specify 
which field bases are referred to.  We will assume that for the
leptons and down-type quarks, it is the mass eigenbasis.  This implies
that up-type quarks have couplings that are rotated by the CKM matrix:
\be
	\tilde\lambda_i \bar Q_i \to
	\tilde \lambda_j\left(\begin{array}{cc} \bar u_{\L,i}
	V_{ij}, &
	\bar d_{\L,j} \end{array}\right)
	\equiv \left(\begin{array}{cc} \tilde\lambda'_i \bar u_i,
	 \tilde\lambda_i \bar d_i \end{array}\right)
\label{CKMmix}
\ee

\begin{table}[b]
\begin{tabular}{|c|c||c|c|}
\hline
  operator & coefficient & operator & coefficient \\
\hline
$(\bar s b)(\bar \mu \mu)$
& $2\tilde\lambda_2\tilde\lambda_3^* |\lambda_2|^2 f_1(r)$ & 
 $(\bar \mu \mu)(\bar\mu \tau)$ & 
$4\lambda_2^*|\lambda_2|^2\lambda_3 f_2(r)$\\
\hline
$(\bar s b)(\bar s b)$ &$\tilde\lambda_2^2 \tilde \lambda_3^{*2}$
& $(\bar d d [\bar u u])(\bar S S)$& 
$ 2 |\lambda_2|^2 |\tilde\lambda_1^{[\prime]}|^2f_1(0)$\\
\hline

\hline
\end{tabular}
\caption{Effective Hamiltonian dimension 6 operators and coefficients; $(\bar f_1 f_2)(\bar
f_3 f_4)$ denotes $(\bar f_{1L}\gamma^\mu f_{2L})(\bar
f_{3L}\gamma_\mu f_{4L})$ (with the exception of $(\bar S S)$, 
which corresponds to $\sfrac12(\bar S \gamma_\mu \gamma_5 S)$) and coefficients are in units of 
$1/(384\pi^2M^2)$ with $m_\Psi = m_\phi = M$ and loop functions
$f_i$ given in text.  $r\equiv m_S^2/M^2$.}
\label{tab2}
\end{table}

The box diagrams relevant for $b\to s \ell^+\ell^-$, $\ell_i\to
3\ell_j$, neutral meson mixing and DM scattering on nucleons
are shown in fig.\ \ref{box}.\footnote{The $SU(2)_L$ charges of the fields in this theory do not allow it to contribute to $b\to s \nu \bar \nu$ at one loop.} 
Evaluating them we find the
effective dimension-6 operators of the same form as (\ref{dim6op}) but
different external states.  The operator coefficients are shown in 
table \ref{tab2}, 
where for simplicity we take $m_\Psi = m_\phi = M$.  Below we will see
that $M\gtrsim 1$\,TeV to meet LHC constraints, but 
 $S$ can be light since it
is dark matter.  The loop functions $f_{1,2}$ are given by
$f_1(r)= (3/2) (3 r^2-2 r^2 \ln (r)-4 r+1)/(1-r)^3$ and
$f_2(r) = 3(-r^2+2 r \ln (r)+1)/(1-r)^3$, normalized such
that $f_{1,2}(1) = 1$ and
$f_1(0) = 3/2$ and $f_2(0) = 3$.

{\bf Flavor constraints.}
To match the observed $B$ anomalies, we require that 
${\tilde\lambda_2\tilde\lambda_3^*|\lambda_2|^2}
	\cong {\left({M / 0.88\,{\rm TeV}}\right)^2}$ \cite{DAmico:2017mtc}.
Therefore the couplings must be of order unity, since LHC searches
discussed below require $M\gtrsim 1\,$TeV.
On the other hand, strong $B_s$ mixing constraints, as determined by
the mass splitting between $B_s$ and $\bar B_s$, limit the 
coefficient of $(\bar s b)^2$ in table \ref{tab2} to be less than 
$1/(408\,{\rm TeV})^2$ at 95\% confidence level (c.l.) 
\cite{Arnan:2016cpy}, giving
the bound
$|\tilde\lambda_2\tilde\lambda_3| \lesssim {M/(6.6\,{\rm TeV}})$.
Combined with the previous determination, this demands large $\lambda_2$,
\be
	|\lambda_2| > 2.9\,\left(M/{\rm TeV}\right)^{1/2} \, .
\label{l2const}
\ee
Analogous bounds arise
from  $K$, $D$ and $B_d$ \cite{Bona:2007vi,Bona:2016bvr} mixing:
$|\tilde\lambda_1\tilde\lambda_2|\lesssim M/(345\,{\rm TeV})$,
$|\tilde\lambda_1'\tilde\lambda_2'|\lesssim M/(110\,{\rm TeV})$.
$|\tilde\lambda_1\tilde\lambda_3|\lesssim M/(17\,{\rm TeV})$.

As an example, suppose that $M = 1\,$TeV and the bound on $B_s$
mixing is saturated.  We can satisfy all other constraints with
hierarchical quark couplings
\be
	|\tilde\lambda_1| = 0.014,\quad
	|\tilde\lambda_2| = 0.14,\quad
	|\tilde\lambda_3| = 1.1,\quad
	|\lambda_2| = 2.9
\label{params}
\ee
If all of the couplings are positive and real,
$\tilde\lambda_1'\tilde\lambda_2' = 0.009$, right at the $D$
mixing 95\% c.l.\ limit.  If $\tilde\lambda_1$ has the opposite
sign to $\tilde\lambda_{2,3}$,
$\tilde\lambda_1'\tilde\lambda_2'$ is smaller,
$\cong 0.004$.

The hierarchical nature of the quark couplings is preserved under
renormalization group running, since they are multiplicatively
renormalized.  The one-loop beta functions take the form
\cite{Luo:2002ey,Machacek:1983tz}
\be
	\beta(\tilde\lambda_i) \equiv \mu {d\over d\mu}\tilde\lambda_i
 = {3\over 16\pi^2}\,\tilde\lambda_i
	\left(\sfrac12|\tilde\lambda_i|^2 + \sum_k |\tilde\lambda_k^2|
	\right)
\ee
For the choice of couplings in (\ref{params}), this leads to a Landau
pole in $\tilde\lambda_2$ at a scale of around $8 m_\phi$, indicating
the need for further new physics at such scales.  For example a
spontaneously broken nonabelian gauge symmetry, such as we already
suggested for explaining the global symmetries of the model, could
avert the Landau pole.

It is technically natural to assume the other leptonic couplings
$\lambda_{1,3}$ are negligible, since they are generated radiatively
only through neutrino mass insertions.  However aesthetically it
may seem peculiar to have $\lambda_2 \gg\lambda_3$.  If 
$\lambda_{1,3}\neq 0$, the box diagrams
leads to lepton
flavor-violating decays such as $\tau\to 3\mu$ and 
$\mu\to 3e$.  However because of the Majorana nature of $S$,
there are crossed box diagrams, shown in fig.\ \ref{box}, that
exactly cancel the uncrossed ones in the limit where external momenta
are neglected in the loop.  
Their amplitudes then scale as $\lambda_3\lambda_2^3 m_\tau^2/m_\phi^4$
and $\lambda_2\lambda_1^3m_\mu^2/m_\phi^4$ respectively.
After comparing them to those of leptonic decays in the SM,
$2\sqrt{2}\,G_F 
(\bar\nu_i\gamma^\mu\ell_i)(\bar\ell_j\gamma^\mu\nu_j)$,
and imposing the experimental limits on the forbidden decay modes 
\cite{Patrignani:2016xqp} we find no significant 
constraints on $\lambda_1$ or $\lambda_3$.

%
\begin{figure}[b]
\hspace{-0.4cm}
\centerline{
\includegraphics[width=\hsize]{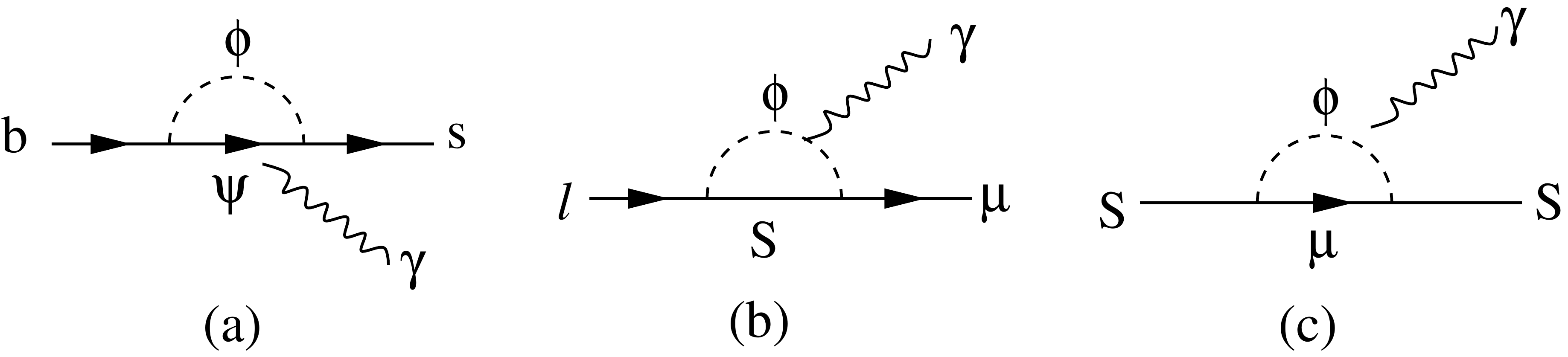}}
\caption{Diagrams leading to (a) $b\to s\gamma$, (b)
 $\tau\to\mu\gamma$, $\mu\to e\gamma$ or $(g-2)_\mu$, 
and (c) $S$ anapole moment.}
\label{b2sg}
\end{figure}

Radiative transitions are another flavor-sensitive observable,
as shown in  
fig.\ \ref{b2sg}.
For $b\to s\gamma$, fig.\ \ref{b2sg}(a) generates the dipole
operator
\be
	{\tilde\lambda^*_3\tilde\lambda_2 e m_b\over 
	32\pi^2}\left( q_\psi{f(R)\over m_\phi^2} -q_\phi {f(R^{-1})\over
	m_\psi^2}\right)(\bar s_\L \slashed{q}\gamma^\mu b_\R)
\label{b2sg_amp}
\ee 
where
$f(R) = (R^3- 6R^2 + 3R + 6 R\ln R + 2)/(6(R-1)^4)$,
$R = m_\Psi^2/m_\phi^2$, $q$ is the photon momentum
and $f(1) = 1/12$. The electric charges $q_i$ of $\Psi$ and $\phi$
are as in table \ref{tab1}.  Due to operator mixing, the
chromomagnetic moment also contributes.  Using the results of 
ref.\ \cite{Arnan:2016cpy},
the Wilson coefficients for our benchmark model with $m_\phi = m_\Psi = 1$ TeV give 
$C_7 + 0.24\, C_8 = -9\times 10^{-3}$, a factor of 10 below the current
limit on this combination from measurements of the branching ratio of $b \to s \gamma$.

Fig.\ \ref{b2sg}(b) gives a contribution to the  anomalous magnetic
moment of the muon of $\Delta (g-2)/2 \cong -(\lambda_2 m_\mu/ \sqrt{96}\pi
m_\phi)^2\cong - 1\times 10^{-10}$, by saturating 
(\ref{l2const}) and taking $m_\phi = 1\,$TeV.  Ultimately this model increases the tension between the measured and predicted values of $g-2$, but the effect is
minimal, 20 times smaller than the SM  discrepancy \cite{Patrignani:2016xqp}. A similar diagram with the photon replaced by the $Z$ leads to a correction of the coupling of the $Z$ to left-handed muons of the form $\delta g_L / g_L^{\rm SM} (q^2=m_Z^2) \cong -(\lambda_2 m_Z / 24 \pi m_\phi)^2 \cong -0.0012\%$ \cite{Arnan:2016cpy}. This is significantly smaller than the uncertainty on the most accurate measurements of this coupling by LEP, $g_L(m_Z^2) = -0.2689 \pm 0.0011$ \cite{ALEPH:2005ab}, which has a 0.4\% error at the $1 \sigma$ level.

If the couplings $\lambda_1$,
$\lambda_3$ are nonzero, there are contributions to 
 $\tau\to\mu\gamma$, $\tau\to e\gamma$, and  $\mu\to e\gamma$,
with partial width $\delta\Gamma \cong \mu_{i,j}^2 m_i^3/8\pi$
\cite{Giunti:2014ixa} where $\mu_{i,j} \cong e\lambda_i\lambda_j m_i / 192\pi^2
m_\phi^2$.
 Using $\lambda_2 = 2.9$ and $m_\phi = 1 \,$TeV, 
the requirement that the partial width of $\tau \to \mu \gamma$ 
induced by the new physics contributions not exceed the measured value
requires $|\lambda_3| < 0.8$, while  $\mu \to e \gamma$ leads to the strong limit
$|\lambda_1| < 1 \times 10^{-3}$.

\begin{figure}
\hspace{-0.4cm}
\centerline{
\includegraphics[width=0.99\hsize]{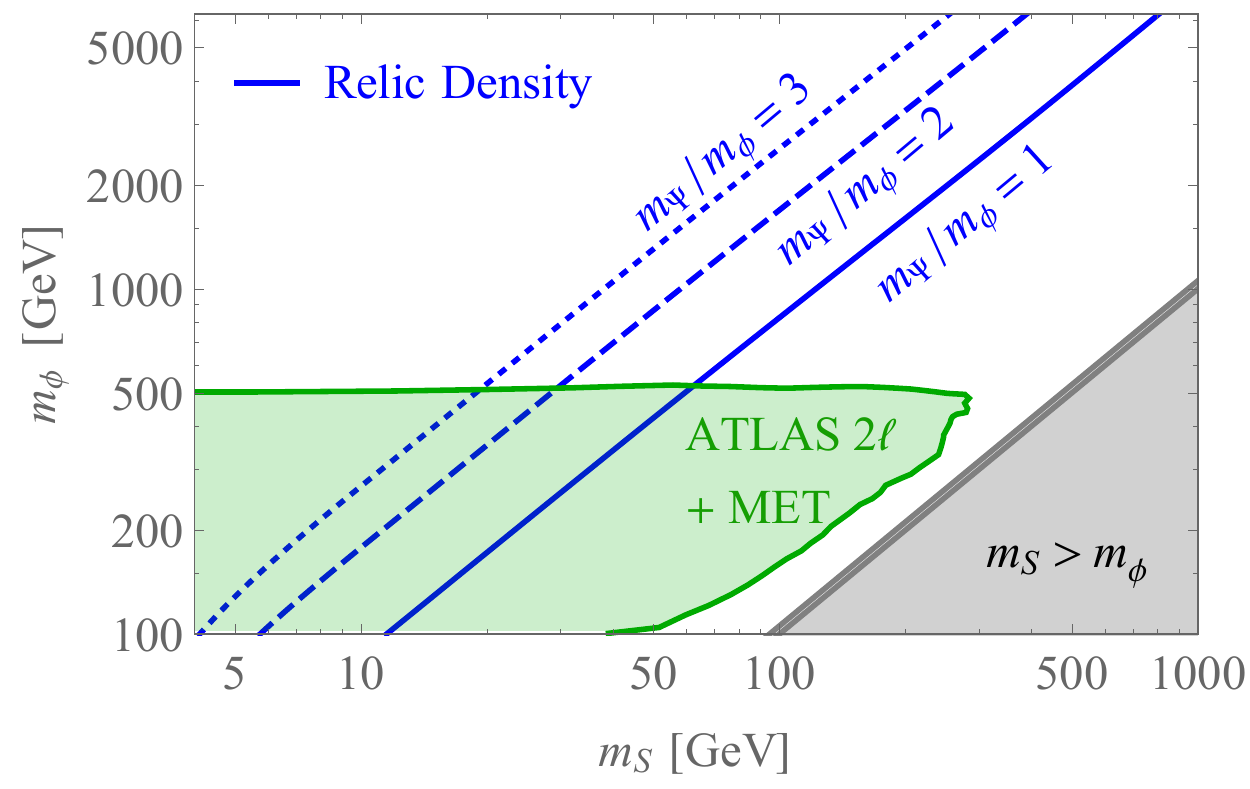}}
\caption{Excluded regions in the plane of $m_\phi$ 
versus $m_S$ from an
ATLAS slepton search \cite{ATLAS:2017uun} (green), and the requirement that $S$ is the lightest particle
so that it can be the DM (grey). The blue lines correspond to values of $m_\phi$ and $m_S$ that give the correct relic density for different values of the ratio  $m_\Psi/m_\phi$.
$\lambda_2$ is set everywhere to the minimum value that allows for
explanation of the flavor anomalies while avoiding 
$B_s$ mixing constraints.}
\label{rd_dwarf}
\end{figure}

\begin{figure}[t]
\hspace{-0.4cm}
\centerline{
\includegraphics[width=0.95\hsize]{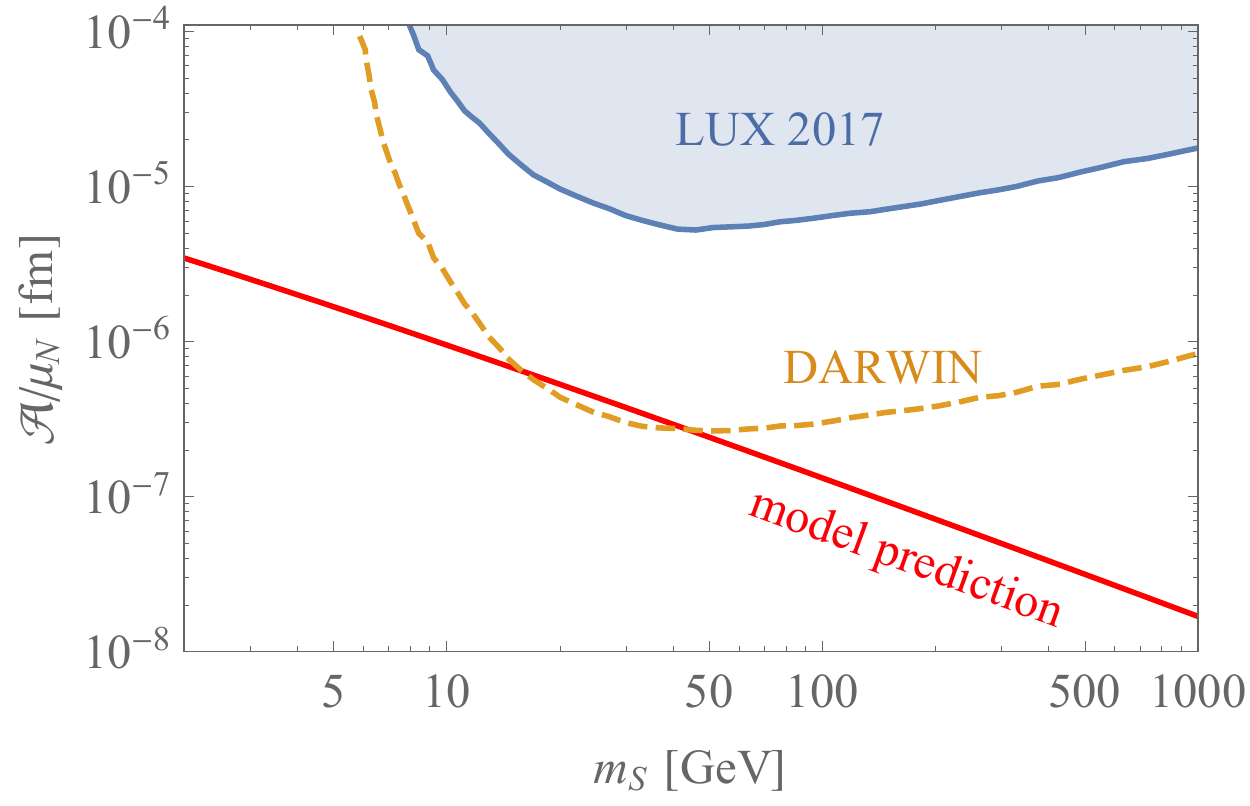}}
\caption{The current limit on the anapole moment from LUX at 90\% c.l. \cite{Kopp:2014tsa,Akerib:2016vxi} and the estimated eventual sensitivity of the DARWIN experiment \cite{Aalbers:2016jon}. The prediction of our model for this quantity, based on the need to achieve the correct relic density and explain the $B$ anomalies, is shown by the red curve.}
\label{anapole}
\end{figure}

{\bf Dark matter constraints.}
The dark matter candidate in our model has tree-level annihilation  to
$\mu\bar\mu$ and $\nu_\mu\bar\nu_\mu$.  The $s$-wave contribution to the cross section is helicity suppressed, so the $v^2$ term dominates \cite{Chang:2014tea}. The total thermally averaged annihilation
cross section, counting both final states, either muons or neutrinos, is
\be
	\langle \sigma v_{\rm rel} \rangle (x) = {|\lambda_2|^4\, m_S^2 (m_\phi^4 + m_S^4) \over
4\pi\,(m_\phi^2+m_S^2)^4 x}
\ee
where $x = m_S / T$. 
To get the
observed relic density \cite{Ade:2015xua}, at the freeze-out temperature $T_f$ this should be roughly equal to the standard value
$\langle\sigma v_{\rm rel}\rangle_0 \cong 4.6\times 10^{-26}{\rm
cm}^3$/s \cite{Steigman:2012nb} appropriate for $p$-wave annihilating Majorana dark matter in the
mass range $m_S\gtrsim 50\,$GeV, that we will see is required
by collider constraints. 
By assuming 
that $\lambda_2$ saturates the inequality (\ref{l2const}) so that
it is no larger than needed to satisfy the flavor
constraints, the relation $\sigma v_{\rm rel} (x_f) = \langle\sigma v_{\rm
rel}\rangle_0$ requires
\be
	m_S = 0.026 \sqrt{x_f} \, m_\phi \, .
\label{msmphi}
\ee
This is valid if $m_\phi \ge m_\Psi$; one can show that (\ref{msmphi}) is further reduced by the factor
$m_\phi/m_\Psi$ if $m_\phi < m_\Psi$.

We verified the previous estimate by numerically solving the Boltzmann
equation with micrOMEGAs 4.3.5 \cite{Belanger:2006is}; contours
corresponding to the cosmologically preferred value $\Omega h^2 =
0.1199$ \cite{Ade:2015xua} are displayed in fig.\ \ref{rd_dwarf}. $S$
annihilations can lead to indirect signals in gamma rays and charged
cosmic rays, but the $p$-wave suppression of  the cross section makes
the limits from such searches very weak. Collider limits are far more
constraining, notably ATLAS searches for 2 leptons and missing
transverse energy \cite{ATLAS:2017uun},  which exclude the green
region in fig.\ \ref{rd_dwarf}.

Because $S$ is a Majorana particle, 
the box diagram for scattering of $S$
off quarks leads only to spin-dependent or velocity-suppressed 
scattering off nucleons. The spin-dependent cross
section for DM scattering off a single nucleon is given by  $\sigma =
\sigma_0 \left( |\tilde \lambda_1|^2 \Delta^{(n)}_d +  |\tilde
\lambda_1'|^2 \Delta^{(n)}_u + |\tilde \lambda_2|^2 \Delta^{(n)}_s
\right)^2,$ where $\sigma_0 = {3 \mu^2_{n,S} |\lambda_2|^4}/{(256
\pi^{5/2} M^2)^2}$ for low-energy scattering (e.g.
\cite{Belanger:2008sj}). The determination of the $\Delta^{(n)}_q$
parameters is reviewed in \cite{Workgroup:2017lvb}. For our benchmark
model with $M = 1$ TeV this leads to $\sigma \sim 10^{-50}\,$cm$^2$
for scattering off neutrons, far below current experimental
limits on spin-dependent scattering from the PICO-60 direct detection
experiment \cite{Amole:2017dex}.

Had the dark matter been Dirac,  diagram (c) of fig.\ \ref{b2sg} would
give both a magnetic moment for  the dark matter  $\mu_S \cong
{e|\lambda_2|^2  m_S/(64\pi^2 m_\phi^2})$,  [approximating $m_S\ll
m_\phi$ consistently with eq.\ (\ref{msmphi})], and a charge-radius
interaction $(\bar S\gamma_\mu S) \partial_\nu F^{\mu\nu}$  that lead
to scattering on protons.  Although the former is below current direct
detection limits, the latter is far too large, which obliges us to 
take $S$ to be Majorana.\footnote{We thank S.\ Okawa for pointing out
the importance of the charge radius contribution.}\ \  Then there is
only an anapole moment  ${\cal A}(\bar S\gamma_\mu\gamma_5 S)\,
\partial_\nu F^{\mu\nu}$, which has been computed and constrained
(using 2013 LUX results) for our class of models in  ref.\
\cite{Kopp:2014tsa}.  We rescale their limit on ${\cal A}$ to reflect
more recent results from  LUX \cite{Akerib:2016vxi}, as well as the
projected eventual sensitivity of DARWIN \cite{Aalbers:2016jon}, in fig.\
\ref{anapole}.  The predicted value is also shown, using
(\ref{l2const}) and  (\ref{msmphi}) with $x_f = 22$ to eliminate
$\lambda_2$ and $m_\phi$ in favor of $m_S$.   For the lowest allowed
value of  $m_S = 60\,$GeV (considering that $m_\phi\gtrsim 500\,$GeV
from LHC constraints), the limit is a factor of $22.5$ weaker than the
prediction, corresponding to a factor of 500 in the cross section. 
This is below the reach of the LZ experiment \cite{Szydagis:2016few},
but slightly above the expected sensitivity of DARWIN, leaving open
the possibility of direct detection.

%
\begin{figure}[t]
\hspace{-0.4cm}
\centerline{
\includegraphics[width=0.95\hsize]{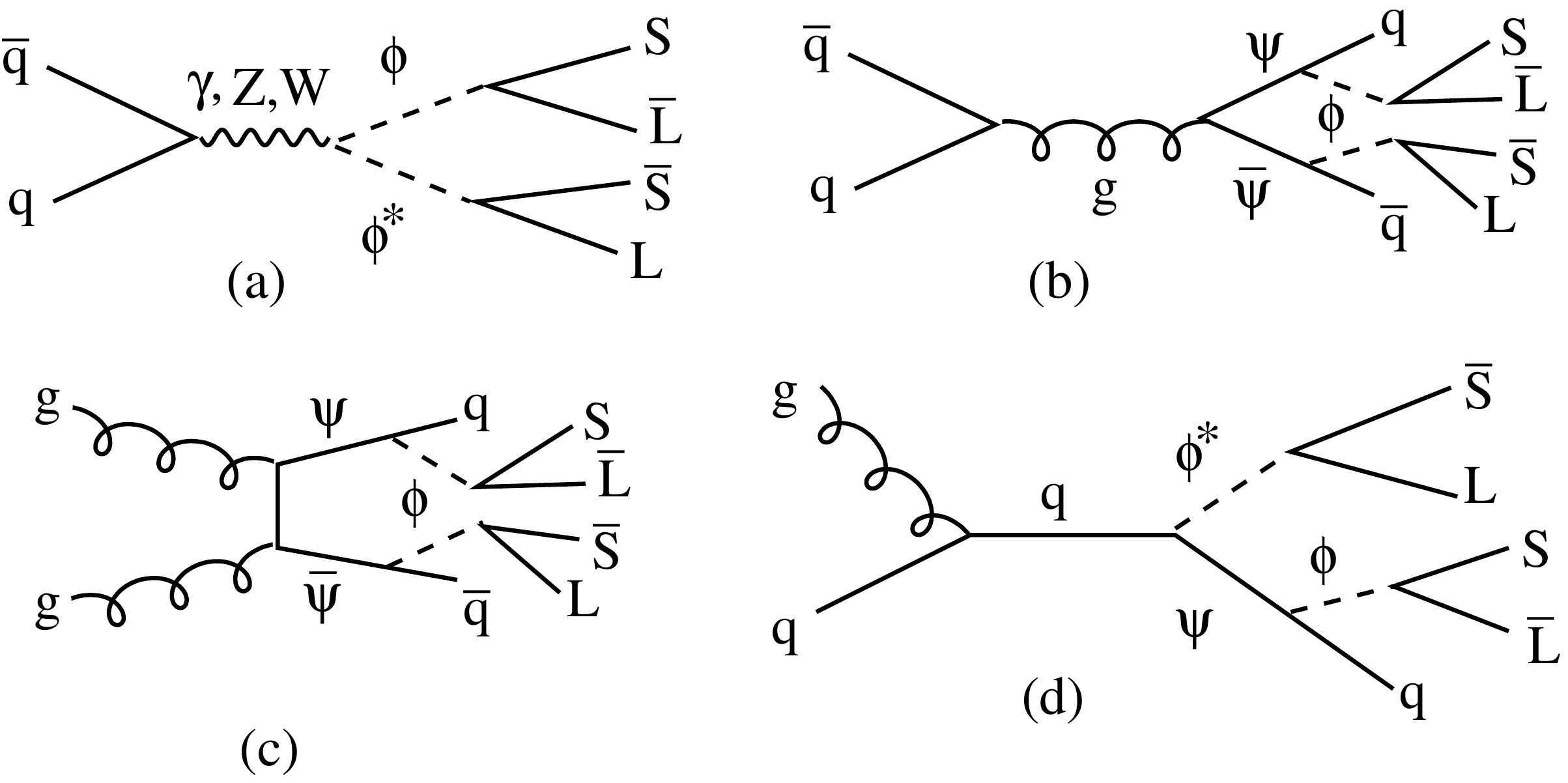}}
\caption{Processes for production of quark jets, leptons, and missing energy.}
\label{lhc}
\end{figure}
%

%
\begin{figure}
\hspace{-0.4cm}
\centerline{
\includegraphics[width=\hsize]{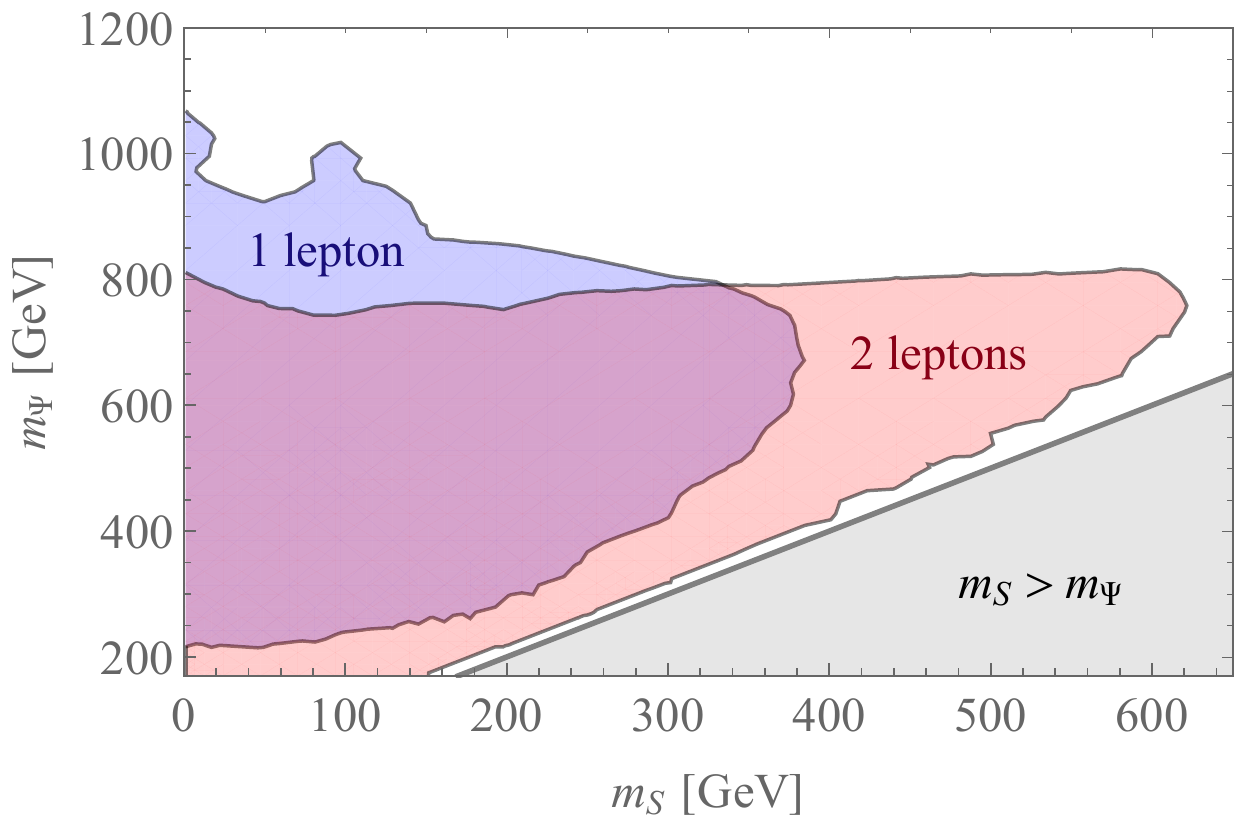}}
\caption{Shaded regions in the $m_S$-$m_\Psi$ plane are excluded
at 95\% c.l. by ATLAS run 2 searches for one (blue) or two (red) leptons, 
jets, and missing energy \cite{ATLAS:2016ljb,ATLAS:2016xcm}. 
For each point, $m_S$ and the couplings are set as described in text 
to satisfy 
flavor and DM relic density constraints.}
\label{collider}
\end{figure}
%

{\bf Collider constraints.} The new states $\phi$ and $\Psi$ carry SM
quantum numbers, and can therefore be pair-produced in particle
collisions.  Fig.\ \ref{lhc} shows the main production modes at a
hadron collider and their decays.  The final states necessarily
include hard lepton pairs, since the splitting between 
$m_\phi$ and $m_S$ must be large, eq.\ (\ref{msmphi}).
This also produces missing energy as the decay products
inevitably include dark matter $S\bar S$ pairs.  Moreover 
hadronic jets appear if $\Psi$ is produced, since $\Psi$ decays into
$\phi$ plus quarks.

For Drell-Yan production of $\phi$-$\phi^*$ pairs, the signal  is
lepton pairs and missing energy, with no jets.   (One of the leptons
is a neutrino if $q\bar q\to W\to \phi^\pm\phi^0$ occurs).  This is
the same final state as in production of slepton pairs, so SUSY
searches \cite{ATLAS:2017uun} may be applied.\footnote{These limits
assume annihilation to all flavors of
both right and left handed sleptons, taken to be degenerate.
Comparing production cross sections of all sleptons to that
of a $\phi \phi^*$ pair using MadGraph \cite{Alwall:2014hca}
indicates that they may be overly stringent for our model;
at 13 TeV, the slepton production cross
section is $\sigma = 1.40\,$fb for $m_{\tilde\ell}= 500$\,GeV, 
whereas
$\sigma = 0.33$\,fb for  $\phi \phi^*$ production with
$m_\phi = 500$\,GeV.} The excluded region is shown in fig.\ \ref{rd_dwarf},
constraining $m_\phi \gtrsim 500\,$GeV for all $m_S$ for which the
relic density can be accommodated.

In diagrams \ref{lhc}(b,c,d), $\Psi$ is produced,
which subsequently decays to $b \mu^+ S$ or $t \bar \nu_\mu S$. Such
final states have been searched for by ATLAS  in 13.3
fb$^{-1}$ of $\sqrt{s} = 13\,$TeV data, including events with one or two leptons, 
jets and missing transverse momentum 
\cite{ATLAS:2016ljb,ATLAS:2016xcm}. These
analyses has been implemented in CheckMATE 2.0.14 \cite{Drees:2013wra},
which we used  to constrain our model, in conjunction with FeynRules
2.3 \cite{Alloul:2013bka} and MadGraph 2.6.0 \cite{Alwall:2014hca}.
20,000 events per model point were generated for the process 
$p p \to \Psi \bar \Psi$ ($p p \to \Psi \phi^*$ is suppressed by the small couplings of $\Psi$ and $\phi$ to first 
generation quarks, or the parton distribution function of $b$ or $t$).
The subsequent showering and hadronisation of the final state partons
was modelled with Pythia 8.230 \cite{Sjostrand:2014zea} and detector simulation was done with Delphes 3.4.1 \cite{deFavereau:2013fsa}.

Fig.~\ref{collider} shows the resulting 95\% c.l.\ limits on $m_\Psi$
versus $m_S$ for models which both explain the flavor anomalies and
give the correct DM relic density. Here $m_\phi$ is set by
eq.~\ref{msmphi} with $x_f = 22$ and the couplings are scaled relative
to (\ref{params}) by the factor $(M/1\,{\rm TeV})^{1/2}$, where $M =
{\rm max}(m_\phi,m_\Psi)$; this choice keeps all the box diagrams 
approximately constant.  At values of $m_S \lesssim 60$ GeV, 
the lowest
values that allow for the correct relic density while avoiding slepton
search constraints, the one-lepton search limits 
$m_\Psi \gtrsim 950$
GeV, except for a narrow window with $m_S$ just below $m_\phi$.  The two-lepton search does not constrain $m_\Psi$ as strongly but
is more sensitive to larger DM masses.

{\bf Conclusions.}  The indications from LHCb of lepton flavor
universality breaking down are currently our best hint of physics
beyond the standard model from colliders.  These anomalies should be 
verified within a few years by further data from LHCb and Belle II
\cite{Albrecht:2017odf}.  If confirmed,  it is not unreasonable to 
expect that the relevant new physics could also shed light on other
shortcomings of the standard model.  We have shown how a very
economical model, in which dark matter plays an essential role, could
be the source of $R(K^{(*)})$ anomalies, while predicting  imminent
tensions in other flavor observables, notably $B_s$ mixing.   The model
may be tested by the next generation of direct detection searches and
can be discovered at the LHC via searches for 
leptons, jets and missing energy.

{\bf Acknowledgment.}  
We thank J.\ Martin Camalich and D.\ London for helpful discussions,
and S. Okawa for alerting us to a problem with the first version
of this paper.  
Our work is supported by the Natural Sciences
and Engineering Research Council (NSERC) of Canada.

\bibliography{loop,leptoquark-ref,Zprime-ref}

\begin{thebibliography}{76}%
\makeatletter
\providecommand \@ifxundefined [1]{%
 \@ifx{#1\undefined}
}%
\providecommand \@ifnum [1]{%
 \ifnum #1\expandafter \@firstoftwo
 \else \expandafter \@secondoftwo
 \fi
}%
\providecommand \@ifx [1]{%
 \ifx #1\expandafter \@firstoftwo
 \else \expandafter \@secondoftwo
 \fi
}%
\providecommand \natexlab [1]{#1}%
\providecommand \enquote  [1]{``#1''}%
\providecommand \bibnamefont  [1]{#1}%
\providecommand \bibfnamefont [1]{#1}%
\providecommand \citenamefont [1]{#1}%
\providecommand \href@noop [0]{\@secondoftwo}%
\providecommand \href [0]{\begingroup \@sanitize@url \@href}%
\providecommand \@href[1]{\@@startlink{#1}\@@href}%
\providecommand \@@href[1]{\endgroup#1\@@endlink}%
\providecommand \@sanitize@url [0]{\catcode `\\12\catcode `\$12\catcode
  `\&12\catcode `\#12\catcode `\^12\catcode `\_12\catcode `\%12\relax}%
\providecommand \@@startlink[1]{}%
\providecommand \@@endlink[0]{}%
\providecommand \url  [0]{\begingroup\@sanitize@url \@url }%
\providecommand \@url [1]{\endgroup\@href {#1}{\urlprefix }}%
\providecommand \urlprefix  [0]{URL }%
\providecommand \Eprint [0]{\href }%
\providecommand \doibase [0]{http://dx.doi.org/}%
\providecommand \selectlanguage [0]{\@gobble}%
\providecommand \bibinfo  [0]{\@secondoftwo}%
\providecommand \bibfield  [0]{\@secondoftwo}%
\providecommand \translation [1]{[#1]}%
\providecommand \BibitemOpen [0]{}%
\providecommand \bibitemStop [0]{}%
\providecommand \bibitemNoStop [0]{.\EOS\space}%
\providecommand \EOS [0]{\spacefactor3000\relax}%
\providecommand \BibitemShut  [1]{\csname bibitem#1\endcsname}%
\let\auto@bib@innerbib\@empty
\bibitem [{\citenamefont {Aaij}\ \emph {et~al.}(2014)\citenamefont {Aaij} \emph
  {et~al.}}]{Aaij:2014ora}%
  \BibitemOpen
  \bibfield  {author} {\bibinfo {author} {\bibfnamefont {Roel}\ \bibnamefont
  {Aaij}} \emph {et~al.} (\bibinfo {collaboration} {LHCb}),\ }\bibfield
  {title} {\enquote {\bibinfo {title} {{Test of lepton universality using
  $B^{+}\rightarrow K^{+}\ell^{+}\ell^{-}$ decays}},}\ }\href {\doibase
  10.1103/PhysRevLett.113.151601} {\bibfield  {journal} {\bibinfo  {journal}
  {Phys. Rev. Lett.}\ }\textbf {\bibinfo {volume} {113}},\ \bibinfo {pages}
  {151601} (\bibinfo {year} {2014})},\ \Eprint {http://arxiv.org/abs/1406.6482}
  {arXiv:1406.6482 [hep-ex]} \BibitemShut {NoStop}%
\bibitem [{\citenamefont {Aaij}\ \emph {et~al.}(2017)\citenamefont {Aaij} \emph
  {et~al.}}]{Aaij:2017vbb}%
  \BibitemOpen
  \bibfield  {author} {\bibinfo {author} {\bibfnamefont {R.}~\bibnamefont
  {Aaij}} \emph {et~al.} (\bibinfo {collaboration} {LHCb}),\ }\bibfield
  {title} {\enquote {\bibinfo {title} {{Test of lepton universality with $B^{0}
  \rightarrow K^{*0}\ell^{+}\ell^{-}$ decays}},}\ }\href {\doibase
  10.1007/JHEP08(2017)055} {\bibfield  {journal} {\bibinfo  {journal} {JHEP}\
  }\textbf {\bibinfo {volume} {08}},\ \bibinfo {pages} {055} (\bibinfo {year}
  {2017})},\ \Eprint {http://arxiv.org/abs/1705.05802} {arXiv:1705.05802
  [hep-ex]} \BibitemShut {NoStop}%
\bibitem [{\citenamefont {Hiller}\ and\ \citenamefont
  {Kruger}(2004)}]{Hiller:2003js}%
  \BibitemOpen
  \bibfield  {author} {\bibinfo {author} {\bibfnamefont {Gudrun}\ \bibnamefont
  {Hiller}}\ and\ \bibinfo {author} {\bibfnamefont {Frank}\ \bibnamefont
  {Kruger}},\ }\bibfield  {title} {\enquote {\bibinfo {title} {{More
  model-independent analysis of $b \to s$ processes}},}\ }\href {\doibase
  10.1103/PhysRevD.69.074020} {\bibfield  {journal} {\bibinfo  {journal} {Phys.
  Rev.}\ }\textbf {\bibinfo {volume} {D69}},\ \bibinfo {pages} {074020}
  (\bibinfo {year} {2004})},\ \Eprint {http://arxiv.org/abs/hep-ph/0310219}
  {arXiv:hep-ph/0310219 [hep-ph]} \BibitemShut {NoStop}%
\bibitem [{\citenamefont {Bordone}\ \emph {et~al.}(2016)\citenamefont
  {Bordone}, \citenamefont {Isidori},\ and\ \citenamefont
  {Pattori}}]{Bordone:2016gaq}%
  \BibitemOpen
  \bibfield  {author} {\bibinfo {author} {\bibfnamefont {Marzia}\ \bibnamefont
  {Bordone}}, \bibinfo {author} {\bibfnamefont {Gino}\ \bibnamefont {Isidori}},
  \ and\ \bibinfo {author} {\bibfnamefont {Andrea}\ \bibnamefont {Pattori}},\
  }\bibfield  {title} {\enquote {\bibinfo {title} {{On the Standard Model
  predictions for $R_K$ and $R_{K^*}$}},}\ }\href {\doibase
  10.1140/epjc/s10052-016-4274-7} {\bibfield  {journal} {\bibinfo  {journal}
  {Eur. Phys. J.}\ }\textbf {\bibinfo {volume} {C76}},\ \bibinfo {pages} {440}
  (\bibinfo {year} {2016})},\ \Eprint {http://arxiv.org/abs/1605.07633}
  {arXiv:1605.07633 [hep-ph]} \BibitemShut {NoStop}%
\bibitem [{\citenamefont {Capdevila}\ \emph {et~al.}(2017)\citenamefont
  {Capdevila}, \citenamefont {Crivellin}, \citenamefont {Descotes-Genon},
  \citenamefont {Matias},\ and\ \citenamefont {Virto}}]{Capdevila:2017bsm}%
  \BibitemOpen
  \bibfield  {author} {\bibinfo {author} {\bibfnamefont {Bernat}\ \bibnamefont
  {Capdevila}}, \bibinfo {author} {\bibfnamefont {Andreas}\ \bibnamefont
  {Crivellin}}, \bibinfo {author} {\bibfnamefont {S\'ebastien}\ \bibnamefont
  {Descotes-Genon}}, \bibinfo {author} {\bibfnamefont {Joaquim}\ \bibnamefont
  {Matias}}, \ and\ \bibinfo {author} {\bibfnamefont {Javier}\ \bibnamefont
  {Virto}},\ }\bibfield  {title} {\enquote {\bibinfo {title} {{Patterns of New
  Physics in $b\to s\ell^+\ell^-$ transitions in the light of recent data}},}\
  }\href@noop {} {\  (\bibinfo {year} {2017})},\ \Eprint
  {http://arxiv.org/abs/1704.05340} {arXiv:1704.05340 [hep-ph]} \BibitemShut
  {NoStop}%
\bibitem [{\citenamefont {Geng}\ \emph {et~al.}(2017)\citenamefont {Geng},
  \citenamefont {Grinstein}, \citenamefont {J{\"a}ger}, \citenamefont
  {Martin~Camalich}, \citenamefont {Ren},\ and\ \citenamefont
  {Shi}}]{Geng:2017svp}%
  \BibitemOpen
  \bibfield  {author} {\bibinfo {author} {\bibfnamefont {Li-Sheng}\
  \bibnamefont {Geng}}, \bibinfo {author} {\bibfnamefont {Benjam\'in}\
  \bibnamefont {Grinstein}}, \bibinfo {author} {\bibfnamefont {Sebastian}\
  \bibnamefont {J{\"a}ger}}, \bibinfo {author} {\bibfnamefont {Jorge}\
  \bibnamefont {Martin~Camalich}}, \bibinfo {author} {\bibfnamefont {Xiu-Lei}\
  \bibnamefont {Ren}}, \ and\ \bibinfo {author} {\bibfnamefont {Rui-Xiang}\
  \bibnamefont {Shi}},\ }\bibfield  {title} {\enquote {\bibinfo {title}
  {{Towards the discovery of new physics with lepton-universality ratios of
  $b\to s\ell\ell$ decays}},}\ }\href {\doibase 10.1103/PhysRevD.96.093006}
  {\bibfield  {journal} {\bibinfo  {journal} {Phys. Rev.}\ }\textbf {\bibinfo
  {volume} {D96}},\ \bibinfo {pages} {093006} (\bibinfo {year} {2017})},\
  \Eprint {http://arxiv.org/abs/1704.05446} {arXiv:1704.05446 [hep-ph]}
  \BibitemShut {NoStop}%
\bibitem [{\citenamefont {Ciuchini}\ \emph {et~al.}(2017)\citenamefont
  {Ciuchini}, \citenamefont {Coutinho}, \citenamefont {Fedele}, \citenamefont
  {Franco}, \citenamefont {Paul}, \citenamefont {Silvestrini},\ and\
  \citenamefont {Valli}}]{Ciuchini:2017mik}%
  \BibitemOpen
  \bibfield  {author} {\bibinfo {author} {\bibfnamefont {Marco}\ \bibnamefont
  {Ciuchini}}, \bibinfo {author} {\bibfnamefont {Antonio~M.}\ \bibnamefont
  {Coutinho}}, \bibinfo {author} {\bibfnamefont {Marco}\ \bibnamefont
  {Fedele}}, \bibinfo {author} {\bibfnamefont {Enrico}\ \bibnamefont {Franco}},
  \bibinfo {author} {\bibfnamefont {Ayan}\ \bibnamefont {Paul}}, \bibinfo
  {author} {\bibfnamefont {Luca}\ \bibnamefont {Silvestrini}}, \ and\ \bibinfo
  {author} {\bibfnamefont {Mauro}\ \bibnamefont {Valli}},\ }\bibfield  {title}
  {\enquote {\bibinfo {title} {{On Flavourful Easter eggs for New Physics
  hunger and Lepton Flavour Universality violation}},}\ }\href {\doibase
  10.1140/epjc/s10052-017-5270-2} {\bibfield  {journal} {\bibinfo  {journal}
  {Eur. Phys. J.}\ }\textbf {\bibinfo {volume} {C77}},\ \bibinfo {pages} {688}
  (\bibinfo {year} {2017})},\ \Eprint {http://arxiv.org/abs/1704.05447}
  {arXiv:1704.05447 [hep-ph]} \BibitemShut {NoStop}%
\bibitem [{\citenamefont {Hiller}\ and\ \citenamefont
  {Nisandzic}(2017)}]{Hiller:2017bzc}%
  \BibitemOpen
  \bibfield  {author} {\bibinfo {author} {\bibfnamefont {Gudrun}\ \bibnamefont
  {Hiller}}\ and\ \bibinfo {author} {\bibfnamefont {Ivan}\ \bibnamefont
  {Nisandzic}},\ }\bibfield  {title} {\enquote {\bibinfo {title} {{$R_K$ and
  $R_{K^{\ast}}$ beyond the standard model}},}\ }\href {\doibase
  10.1103/PhysRevD.96.035003} {\bibfield  {journal} {\bibinfo  {journal} {Phys.
  Rev.}\ }\textbf {\bibinfo {volume} {D96}},\ \bibinfo {pages} {035003}
  (\bibinfo {year} {2017})},\ \Eprint {http://arxiv.org/abs/1704.05444}
  {arXiv:1704.05444 [hep-ph]} \BibitemShut {NoStop}%
\bibitem [{\citenamefont {Altmannshofer}\ \emph {et~al.}(2017)\citenamefont
  {Altmannshofer}, \citenamefont {Stangl},\ and\ \citenamefont
  {Straub}}]{Altmannshofer:2017yso}%
  \BibitemOpen
  \bibfield  {author} {\bibinfo {author} {\bibfnamefont {Wolfgang}\
  \bibnamefont {Altmannshofer}}, \bibinfo {author} {\bibfnamefont {Peter}\
  \bibnamefont {Stangl}}, \ and\ \bibinfo {author} {\bibfnamefont {David~M.}\
  \bibnamefont {Straub}},\ }\bibfield  {title} {\enquote {\bibinfo {title}
  {{Interpreting Hints for Lepton Flavor Universality Violation}},}\ }\href
  {\doibase 10.1103/PhysRevD.96.055008} {\bibfield  {journal} {\bibinfo
  {journal} {Phys. Rev.}\ }\textbf {\bibinfo {volume} {D96}},\ \bibinfo {pages}
  {055008} (\bibinfo {year} {2017})},\ \Eprint
  {http://arxiv.org/abs/1704.05435} {arXiv:1704.05435 [hep-ph]} \BibitemShut
  {NoStop}%
\bibitem [{\citenamefont {D'Amico}\ \emph {et~al.}(2017)\citenamefont
  {D'Amico}, \citenamefont {Nardecchia}, \citenamefont {Panci}, \citenamefont
  {Sannino}, \citenamefont {Strumia}, \citenamefont {Torre},\ and\
  \citenamefont {Urbano}}]{DAmico:2017mtc}%
  \BibitemOpen
  \bibfield  {author} {\bibinfo {author} {\bibfnamefont {Guido}\ \bibnamefont
  {D'Amico}}, \bibinfo {author} {\bibfnamefont {Marco}\ \bibnamefont
  {Nardecchia}}, \bibinfo {author} {\bibfnamefont {Paolo}\ \bibnamefont
  {Panci}}, \bibinfo {author} {\bibfnamefont {Francesco}\ \bibnamefont
  {Sannino}}, \bibinfo {author} {\bibfnamefont {Alessandro}\ \bibnamefont
  {Strumia}}, \bibinfo {author} {\bibfnamefont {Riccardo}\ \bibnamefont
  {Torre}}, \ and\ \bibinfo {author} {\bibfnamefont {Alfredo}\ \bibnamefont
  {Urbano}},\ }\bibfield  {title} {\enquote {\bibinfo {title} {{Flavour
  anomalies after the $R_{K^*}$ measurement}},}\ }\href {\doibase
  10.1007/JHEP09(2017)010} {\bibfield  {journal} {\bibinfo  {journal} {JHEP}\
  }\textbf {\bibinfo {volume} {09}},\ \bibinfo {pages} {010} (\bibinfo {year}
  {2017})},\ \Eprint {http://arxiv.org/abs/1704.05438} {arXiv:1704.05438
  [hep-ph]} \BibitemShut {NoStop}%
\bibitem [{\citenamefont {Di~Chiara}\ \emph {et~al.}(2017)\citenamefont
  {Di~Chiara}, \citenamefont {Fowlie}, \citenamefont {Fraser}, \citenamefont
  {Marzo}, \citenamefont {Marzola}, \citenamefont {Raidal},\ and\ \citenamefont
  {Spethmann}}]{DiChiara:2017cjq}%
  \BibitemOpen
  \bibfield  {author} {\bibinfo {author} {\bibfnamefont {Stefano}\ \bibnamefont
  {Di~Chiara}}, \bibinfo {author} {\bibfnamefont {Andrew}\ \bibnamefont
  {Fowlie}}, \bibinfo {author} {\bibfnamefont {Sean}\ \bibnamefont {Fraser}},
  \bibinfo {author} {\bibfnamefont {Carlo}\ \bibnamefont {Marzo}}, \bibinfo
  {author} {\bibfnamefont {Luca}\ \bibnamefont {Marzola}}, \bibinfo {author}
  {\bibfnamefont {Martti}\ \bibnamefont {Raidal}}, \ and\ \bibinfo {author}
  {\bibfnamefont {Christian}\ \bibnamefont {Spethmann}},\ }\bibfield  {title}
  {\enquote {\bibinfo {title} {{Minimal flavor-changing $Z'$ models and muon
  $g-2$ after the $R_{K^*}$ measurement}},}\ }\href {\doibase
  10.1016/j.nuclphysb.2017.08.003} {\bibfield  {journal} {\bibinfo  {journal}
  {Nucl. Phys.}\ }\textbf {\bibinfo {volume} {B923}},\ \bibinfo {pages}
  {245--257} (\bibinfo {year} {2017})},\ \Eprint
  {http://arxiv.org/abs/1704.06200} {arXiv:1704.06200 [hep-ph]} \BibitemShut
  {NoStop}%
\bibitem [{\citenamefont {Chiang}\ \emph
  {et~al.}(2017{\natexlab{a}})\citenamefont {Chiang}, \citenamefont {He},
  \citenamefont {Tandean},\ and\ \citenamefont {Yuan}}]{Chiang:2017hlj}%
  \BibitemOpen
  \bibfield  {author} {\bibinfo {author} {\bibfnamefont {Cheng-Wei}\
  \bibnamefont {Chiang}}, \bibinfo {author} {\bibfnamefont {Xiao-Gang}\
  \bibnamefont {He}}, \bibinfo {author} {\bibfnamefont {Jusak}\ \bibnamefont
  {Tandean}}, \ and\ \bibinfo {author} {\bibfnamefont {Xing-Bo}\ \bibnamefont
  {Yuan}},\ }\bibfield  {title} {\enquote {\bibinfo {title} {{$R_{K^{(*)}}$ and
  related $b\to s\ell\bar\ell$ anomalies in minimal flavor violation framework
  with $Z'$ boson}},}\ }\href {\doibase 10.1103/PhysRevD.96.115022} {\bibfield
  {journal} {\bibinfo  {journal} {Phys. Rev.}\ }\textbf {\bibinfo {volume}
  {D96}},\ \bibinfo {pages} {115022} (\bibinfo {year} {2017}{\natexlab{a}})},\
  \Eprint {http://arxiv.org/abs/1706.02696} {arXiv:1706.02696 [hep-ph]}
  \BibitemShut {NoStop}%
\bibitem [{\citenamefont {Dalchenko}\ \emph {et~al.}(2017)\citenamefont
  {Dalchenko}, \citenamefont {Dutta}, \citenamefont {Eusebi}, \citenamefont
  {Huang}, \citenamefont {Kamon},\ and\ \citenamefont
  {Rathjens}}]{Dalchenko:2017shg}%
  \BibitemOpen
  \bibfield  {author} {\bibinfo {author} {\bibfnamefont {Mykhailo}\
  \bibnamefont {Dalchenko}}, \bibinfo {author} {\bibfnamefont {Bhaskar}\
  \bibnamefont {Dutta}}, \bibinfo {author} {\bibfnamefont {Ricardo}\
  \bibnamefont {Eusebi}}, \bibinfo {author} {\bibfnamefont {Peisi}\
  \bibnamefont {Huang}}, \bibinfo {author} {\bibfnamefont {Teruki}\
  \bibnamefont {Kamon}}, \ and\ \bibinfo {author} {\bibfnamefont {Denis}\
  \bibnamefont {Rathjens}},\ }\bibfield  {title} {\enquote {\bibinfo {title}
  {{Bottom-quark Fusion Processes at the LHC for Probing $Z^{\prime}$ Models
  and B-meson Decay Anomalies}},}\ }\href@noop {} {\  (\bibinfo {year}
  {2017})},\ \Eprint {http://arxiv.org/abs/1707.07016} {arXiv:1707.07016
  [hep-ph]} \BibitemShut {NoStop}%
\bibitem [{\citenamefont {Agrawal}\ \emph {et~al.}(2018)\citenamefont
  {Agrawal}, \citenamefont {Saha},\ and\ \citenamefont
  {Shivaji}}]{Agrawal:2017cbs}%
  \BibitemOpen
  \bibfield  {author} {\bibinfo {author} {\bibfnamefont {Pankaj}\ \bibnamefont
  {Agrawal}}, \bibinfo {author} {\bibfnamefont {Debashis}\ \bibnamefont
  {Saha}}, \ and\ \bibinfo {author} {\bibfnamefont {Ambresh}\ \bibnamefont
  {Shivaji}},\ }\bibfield  {title} {\enquote {\bibinfo {title} {{Production of
  $HHH$ and $HHV(V=\gamma,Z)$ at the hadron colliders}},}\ }\href {\doibase
  10.1103/PhysRevD.97.036006} {\bibfield  {journal} {\bibinfo  {journal} {Phys.
  Rev.}\ }\textbf {\bibinfo {volume} {D97}},\ \bibinfo {pages} {036006}
  (\bibinfo {year} {2018})},\ \Eprint {http://arxiv.org/abs/1708.03580}
  {arXiv:1708.03580 [hep-ph]} \BibitemShut {NoStop}%
\bibitem [{\citenamefont {Romao}\ \emph {et~al.}(2017)\citenamefont {Romao},
  \citenamefont {King},\ and\ \citenamefont {Leontaris}}]{Romao:2017qnu}%
  \BibitemOpen
  \bibfield  {author} {\bibinfo {author} {\bibfnamefont {Miguel~Crispim}\
  \bibnamefont {Romao}}, \bibinfo {author} {\bibfnamefont {Stephen~F.}\
  \bibnamefont {King}}, \ and\ \bibinfo {author} {\bibfnamefont {George~K.}\
  \bibnamefont {Leontaris}},\ }\bibfield  {title} {\enquote {\bibinfo {title}
  {{Non-universal $Z'$ from Fluxed GUTs}},}\ }\href@noop {} {\  (\bibinfo
  {year} {2017})},\ \Eprint {http://arxiv.org/abs/1710.02349} {arXiv:1710.02349
  [hep-ph]} \BibitemShut {NoStop}%
\bibitem [{\citenamefont {Faisel}\ and\ \citenamefont
  {Tandean}(2018)}]{Faisel:2017glo}%
  \BibitemOpen
  \bibfield  {author} {\bibinfo {author} {\bibfnamefont {Gaber}\ \bibnamefont
  {Faisel}}\ and\ \bibinfo {author} {\bibfnamefont {Jusak}\ \bibnamefont
  {Tandean}},\ }\bibfield  {title} {\enquote {\bibinfo {title} {{Connecting $
  b\to s\ell \overline{\ell} $ anomalies to enhanced rare nonleptonic $
  {\overline{B}}_s^0 $ decays in $Z'$ model}},}\ }\href {\doibase
  10.1007/JHEP02(2018)074} {\bibfield  {journal} {\bibinfo  {journal} {JHEP}\
  }\textbf {\bibinfo {volume} {02}},\ \bibinfo {pages} {074} (\bibinfo {year}
  {2018})},\ \Eprint {http://arxiv.org/abs/1710.11102} {arXiv:1710.11102
  [hep-ph]} \BibitemShut {NoStop}%
\bibitem [{\citenamefont {Falkowski}\ \emph {et~al.}(2018)\citenamefont
  {Falkowski}, \citenamefont {King}, \citenamefont {Perdomo},\ and\
  \citenamefont {Pierre}}]{Falkowski:2018dsl}%
  \BibitemOpen
  \bibfield  {author} {\bibinfo {author} {\bibfnamefont {Adam}\ \bibnamefont
  {Falkowski}}, \bibinfo {author} {\bibfnamefont {Stephen~F.}\ \bibnamefont
  {King}}, \bibinfo {author} {\bibfnamefont {Elena}\ \bibnamefont {Perdomo}}, \
  and\ \bibinfo {author} {\bibfnamefont {Mathias}\ \bibnamefont {Pierre}},\
  }\bibfield  {title} {\enquote {\bibinfo {title} {{Flavourful $Z'$ portal for
  vector-like neutrino Dark Matter and $R_{K^{(*)}}$}},}\ }\href@noop {} {\
  (\bibinfo {year} {2018})},\ \Eprint {http://arxiv.org/abs/1803.04430}
  {arXiv:1803.04430 [hep-ph]} \BibitemShut {NoStop}%
\bibitem [{\citenamefont {Kohda}\ \emph {et~al.}(2018)\citenamefont {Kohda},
  \citenamefont {Modak},\ and\ \citenamefont {Soffer}}]{Kohda:2018xbc}%
  \BibitemOpen
  \bibfield  {author} {\bibinfo {author} {\bibfnamefont {Masaya}\ \bibnamefont
  {Kohda}}, \bibinfo {author} {\bibfnamefont {Tanmoy}\ \bibnamefont {Modak}}, \
  and\ \bibinfo {author} {\bibfnamefont {Abner}\ \bibnamefont {Soffer}},\
  }\bibfield  {title} {\enquote {\bibinfo {title} {{Identifying a $Z'$ behind
  $b \to s \ell \ell$ anomalies at the LHC}},}\ }\href@noop {} {\  (\bibinfo
  {year} {2018})},\ \Eprint {http://arxiv.org/abs/1803.07492} {arXiv:1803.07492
  [hep-ph]} \BibitemShut {NoStop}%
\bibitem [{\citenamefont {Chen}\ \emph {et~al.}(2017)\citenamefont {Chen},
  \citenamefont {Nomura},\ and\ \citenamefont {Okada}}]{Chen:2017hir}%
  \BibitemOpen
  \bibfield  {author} {\bibinfo {author} {\bibfnamefont {Chuan-Hung}\
  \bibnamefont {Chen}}, \bibinfo {author} {\bibfnamefont {Takaaki}\
  \bibnamefont {Nomura}}, \ and\ \bibinfo {author} {\bibfnamefont {Hiroshi}\
  \bibnamefont {Okada}},\ }\bibfield  {title} {\enquote {\bibinfo {title}
  {{Excesses of muon $g-2$, $R_{D^{(\ast)}}$, and $R_K$ in a leptoquark
  model}},}\ }\href {\doibase 10.1016/j.physletb.2017.10.005} {\bibfield
  {journal} {\bibinfo  {journal} {Phys. Lett.}\ }\textbf {\bibinfo {volume}
  {B774}},\ \bibinfo {pages} {456--464} (\bibinfo {year} {2017})},\ \Eprint
  {http://arxiv.org/abs/1703.03251} {arXiv:1703.03251 [hep-ph]} \BibitemShut
  {NoStop}%
\bibitem [{\citenamefont {Crivellin}\ \emph {et~al.}(2017)\citenamefont
  {Crivellin}, \citenamefont {Müller},\ and\ \citenamefont
  {Ota}}]{Crivellin:2017zlb}%
  \BibitemOpen
  \bibfield  {author} {\bibinfo {author} {\bibfnamefont {Andreas}\ \bibnamefont
  {Crivellin}}, \bibinfo {author} {\bibfnamefont {Dario}\ \bibnamefont
  {Müller}}, \ and\ \bibinfo {author} {\bibfnamefont {Toshihiko}\ \bibnamefont
  {Ota}},\ }\bibfield  {title} {\enquote {\bibinfo {title} {{Simultaneous
  explanation of R(D$^{*}$) and $b\to s \mu^+\mu^-$: the last scalar
  leptoquarks standing}},}\ }\href {\doibase 10.1007/JHEP09(2017)040}
  {\bibfield  {journal} {\bibinfo  {journal} {JHEP}\ }\textbf {\bibinfo
  {volume} {09}},\ \bibinfo {pages} {040} (\bibinfo {year} {2017})},\ \Eprint
  {http://arxiv.org/abs/1703.09226} {arXiv:1703.09226 [hep-ph]} \BibitemShut
  {NoStop}%
\bibitem [{\citenamefont {Cai}\ \emph {et~al.}(2017)\citenamefont {Cai},
  \citenamefont {Gargalionis}, \citenamefont {Schmidt},\ and\ \citenamefont
  {Volkas}}]{Cai:2017wry}%
  \BibitemOpen
  \bibfield  {author} {\bibinfo {author} {\bibfnamefont {Yi}~\bibnamefont
  {Cai}}, \bibinfo {author} {\bibfnamefont {John}\ \bibnamefont {Gargalionis}},
  \bibinfo {author} {\bibfnamefont {Michael~A.}\ \bibnamefont {Schmidt}}, \
  and\ \bibinfo {author} {\bibfnamefont {Raymond~R.}\ \bibnamefont {Volkas}},\
  }\bibfield  {title} {\enquote {\bibinfo {title} {{Reconsidering the One
  Leptoquark solution: flavor anomalies and neutrino mass}},}\ }\href {\doibase
  10.1007/JHEP10(2017)047} {\bibfield  {journal} {\bibinfo  {journal} {JHEP}\
  }\textbf {\bibinfo {volume} {10}},\ \bibinfo {pages} {047} (\bibinfo {year}
  {2017})},\ \Eprint {http://arxiv.org/abs/1704.05849} {arXiv:1704.05849
  [hep-ph]} \BibitemShut {NoStop}%
\bibitem [{\citenamefont {Chauhan}\ \emph {et~al.}(2017)\citenamefont
  {Chauhan}, \citenamefont {Kindra},\ and\ \citenamefont
  {Narang}}]{Chauhan:2017ndd}%
  \BibitemOpen
  \bibfield  {author} {\bibinfo {author} {\bibfnamefont {Bhavesh}\ \bibnamefont
  {Chauhan}}, \bibinfo {author} {\bibfnamefont {Bharti}\ \bibnamefont
  {Kindra}}, \ and\ \bibinfo {author} {\bibfnamefont {Ashish}\ \bibnamefont
  {Narang}},\ }\bibfield  {title} {\enquote {\bibinfo {title} {{A Leptoquark
  explanation for $(g-2)_\mu$, $R_K$, $R_{K^*}$ and, IceCube PeV events}},}\
  }\href@noop {} {\  (\bibinfo {year} {2017})},\ \Eprint
  {http://arxiv.org/abs/1706.04598} {arXiv:1706.04598 [hep-ph]} \BibitemShut
  {NoStop}%
\bibitem [{\citenamefont {Diaz}\ \emph {et~al.}(2017)\citenamefont {Diaz},
  \citenamefont {Schmaltz},\ and\ \citenamefont {Zhong}}]{Diaz:2017lit}%
  \BibitemOpen
  \bibfield  {author} {\bibinfo {author} {\bibfnamefont {Bastian}\ \bibnamefont
  {Diaz}}, \bibinfo {author} {\bibfnamefont {Martin}\ \bibnamefont {Schmaltz}},
  \ and\ \bibinfo {author} {\bibfnamefont {Yi-Ming}\ \bibnamefont {Zhong}},\
  }\bibfield  {title} {\enquote {\bibinfo {title} {{The leptoquark Hunter’s
  guide: Pair production}},}\ }\href {\doibase 10.1007/JHEP10(2017)097}
  {\bibfield  {journal} {\bibinfo  {journal} {JHEP}\ }\textbf {\bibinfo
  {volume} {10}},\ \bibinfo {pages} {097} (\bibinfo {year} {2017})},\ \Eprint
  {http://arxiv.org/abs/1706.05033} {arXiv:1706.05033 [hep-ph]} \BibitemShut
  {NoStop}%
\bibitem [{\citenamefont {Doršner}\ \emph {et~al.}(2017)\citenamefont
  {Doršner}, \citenamefont {Fajfer}, \citenamefont {Faroughy},\ and\
  \citenamefont {Košnik}}]{Dorsner:2017ufx}%
  \BibitemOpen
  \bibfield  {author} {\bibinfo {author} {\bibfnamefont {Ilja}\ \bibnamefont
  {Doršner}}, \bibinfo {author} {\bibfnamefont {Svjetlana}\ \bibnamefont
  {Fajfer}}, \bibinfo {author} {\bibfnamefont {Darius~A.}\ \bibnamefont
  {Faroughy}}, \ and\ \bibinfo {author} {\bibfnamefont {Nejc}\ \bibnamefont
  {Košnik}},\ }\bibfield  {title} {\enquote {\bibinfo {title} {{The role of
  the $S_3$ GUT leptoquark in flavor universality and collider searches}},}\
  }\href {\doibase 10.1007/JHEP10(2017)188} {\  (\bibinfo {year} {2017}),\
  10.1007/JHEP10(2017)188},\ \bibinfo {note} {[JHEP10,188(2017)]},\ \Eprint
  {http://arxiv.org/abs/1706.07779} {arXiv:1706.07779 [hep-ph]} \BibitemShut
  {NoStop}%
\bibitem [{\citenamefont {Crivellin}\ \emph {et~al.}(2018)\citenamefont
  {Crivellin}, \citenamefont {Müller}, \citenamefont {Signer},\ and\
  \citenamefont {Ulrich}}]{Crivellin:2017dsk}%
  \BibitemOpen
  \bibfield  {author} {\bibinfo {author} {\bibfnamefont {Andreas}\ \bibnamefont
  {Crivellin}}, \bibinfo {author} {\bibfnamefont {Dario}\ \bibnamefont
  {Müller}}, \bibinfo {author} {\bibfnamefont {Adrian}\ \bibnamefont
  {Signer}}, \ and\ \bibinfo {author} {\bibfnamefont {Yannick}\ \bibnamefont
  {Ulrich}},\ }\bibfield  {title} {\enquote {\bibinfo {title} {{Correlating
  lepton flavor universality violation in $B$ decays with $\mu\to e\gamma$
  using leptoquarks}},}\ }\href {\doibase 10.1103/PhysRevD.97.015019}
  {\bibfield  {journal} {\bibinfo  {journal} {Phys. Rev.}\ }\textbf {\bibinfo
  {volume} {D97}},\ \bibinfo {pages} {015019} (\bibinfo {year} {2018})},\
  \Eprint {http://arxiv.org/abs/1706.08511} {arXiv:1706.08511 [hep-ph]}
  \BibitemShut {NoStop}%
\bibitem [{\citenamefont {Aloni}\ \emph {et~al.}(2017)\citenamefont {Aloni},
  \citenamefont {Dery}, \citenamefont {Frugiuele},\ and\ \citenamefont
  {Nir}}]{Aloni:2017ixa}%
  \BibitemOpen
  \bibfield  {author} {\bibinfo {author} {\bibfnamefont {Daniel}\ \bibnamefont
  {Aloni}}, \bibinfo {author} {\bibfnamefont {Avital}\ \bibnamefont {Dery}},
  \bibinfo {author} {\bibfnamefont {Claudia}\ \bibnamefont {Frugiuele}}, \ and\
  \bibinfo {author} {\bibfnamefont {Yosef}\ \bibnamefont {Nir}},\ }\bibfield
  {title} {\enquote {\bibinfo {title} {{Testing minimal flavor violation in
  leptoquark models of the $ {R_K}_{{}^{\left(\ast \right)}} $ anomaly}},}\
  }\href {\doibase 10.1007/JHEP11(2017)109} {\bibfield  {journal} {\bibinfo
  {journal} {JHEP}\ }\textbf {\bibinfo {volume} {11}},\ \bibinfo {pages} {109}
  (\bibinfo {year} {2017})},\ \Eprint {http://arxiv.org/abs/1708.06161}
  {arXiv:1708.06161 [hep-ph]} \BibitemShut {NoStop}%
\bibitem [{\citenamefont {Assad}\ \emph {et~al.}(2018)\citenamefont {Assad},
  \citenamefont {Fornal},\ and\ \citenamefont {Grinstein}}]{Assad:2017iib}%
  \BibitemOpen
  \bibfield  {author} {\bibinfo {author} {\bibfnamefont {Nima}\ \bibnamefont
  {Assad}}, \bibinfo {author} {\bibfnamefont {Bartosz}\ \bibnamefont {Fornal}},
  \ and\ \bibinfo {author} {\bibfnamefont {Benjamin}\ \bibnamefont
  {Grinstein}},\ }\bibfield  {title} {\enquote {\bibinfo {title} {{Baryon
  Number and Lepton Universality Violation in Leptoquark and Diquark
  Models}},}\ }\href {\doibase 10.1016/j.physletb.2017.12.042} {\bibfield
  {journal} {\bibinfo  {journal} {Phys. Lett.}\ }\textbf {\bibinfo {volume}
  {B777}},\ \bibinfo {pages} {324--331} (\bibinfo {year} {2018})},\ \Eprint
  {http://arxiv.org/abs/1708.06350} {arXiv:1708.06350 [hep-ph]} \BibitemShut
  {NoStop}%
\bibitem [{\citenamefont {Di~Luzio}\ \emph {et~al.}(2017)\citenamefont
  {Di~Luzio}, \citenamefont {Greljo},\ and\ \citenamefont
  {Nardecchia}}]{DiLuzio:2017vat}%
  \BibitemOpen
  \bibfield  {author} {\bibinfo {author} {\bibfnamefont {Luca}\ \bibnamefont
  {Di~Luzio}}, \bibinfo {author} {\bibfnamefont {Admir}\ \bibnamefont
  {Greljo}}, \ and\ \bibinfo {author} {\bibfnamefont {Marco}\ \bibnamefont
  {Nardecchia}},\ }\bibfield  {title} {\enquote {\bibinfo {title} {{Gauge
  leptoquark as the origin of B-physics anomalies}},}\ }\href {\doibase
  10.1103/PhysRevD.96.115011} {\bibfield  {journal} {\bibinfo  {journal} {Phys.
  Rev.}\ }\textbf {\bibinfo {volume} {D96}},\ \bibinfo {pages} {115011}
  (\bibinfo {year} {2017})},\ \Eprint {http://arxiv.org/abs/1708.08450}
  {arXiv:1708.08450 [hep-ph]} \BibitemShut {NoStop}%
\bibitem [{\citenamefont {Calibbi}\ \emph {et~al.}(2017)\citenamefont
  {Calibbi}, \citenamefont {Crivellin},\ and\ \citenamefont
  {Li}}]{Calibbi:2017qbu}%
  \BibitemOpen
  \bibfield  {author} {\bibinfo {author} {\bibfnamefont {Lorenzo}\ \bibnamefont
  {Calibbi}}, \bibinfo {author} {\bibfnamefont {Andreas}\ \bibnamefont
  {Crivellin}}, \ and\ \bibinfo {author} {\bibfnamefont {Tianjun}\ \bibnamefont
  {Li}},\ }\bibfield  {title} {\enquote {\bibinfo {title} {{A model of vector
  leptoquarks in view of the $B$-physics anomalies}},}\ }\href@noop {} {\
  (\bibinfo {year} {2017})},\ \Eprint {http://arxiv.org/abs/1709.00692}
  {arXiv:1709.00692 [hep-ph]} \BibitemShut {NoStop}%
\bibitem [{\citenamefont {Dey}\ \emph {et~al.}(2017)\citenamefont {Dey},
  \citenamefont {Kar}, \citenamefont {Mitra}, \citenamefont {Spannowsky},\ and\
  \citenamefont {Vincent}}]{Dey:2017ede}%
  \BibitemOpen
  \bibfield  {author} {\bibinfo {author} {\bibfnamefont {Ujjal~Kumar}\
  \bibnamefont {Dey}}, \bibinfo {author} {\bibfnamefont {Deepak}\ \bibnamefont
  {Kar}}, \bibinfo {author} {\bibfnamefont {Manimala}\ \bibnamefont {Mitra}},
  \bibinfo {author} {\bibfnamefont {Michael}\ \bibnamefont {Spannowsky}}, \
  and\ \bibinfo {author} {\bibfnamefont {Aaron~C.}\ \bibnamefont {Vincent}},\
  }\bibfield  {title} {\enquote {\bibinfo {title} {{Searching for Leptoquarks
  at IceCube and the LHC}},}\ }\href@noop {} {\  (\bibinfo {year} {2017})},\
  \Eprint {http://arxiv.org/abs/1709.02009} {arXiv:1709.02009 [hep-ph]}
  \BibitemShut {NoStop}%
\bibitem [{\citenamefont {Chauhan}\ and\ \citenamefont
  {Kindra}(2017)}]{Chauhan:2017uil}%
  \BibitemOpen
  \bibfield  {author} {\bibinfo {author} {\bibfnamefont {Bhavesh}\ \bibnamefont
  {Chauhan}}\ and\ \bibinfo {author} {\bibfnamefont {Bharti}\ \bibnamefont
  {Kindra}},\ }\bibfield  {title} {\enquote {\bibinfo {title} {{Invoking Chiral
  Vector Leptoquark to explain LFU violation in B Decays}},}\ }\href@noop {} {\
   (\bibinfo {year} {2017})},\ \Eprint {http://arxiv.org/abs/1709.09989}
  {arXiv:1709.09989 [hep-ph]} \BibitemShut {NoStop}%
\bibitem [{\citenamefont {Müller}(2018)}]{Muller:2018nwq}%
  \BibitemOpen
  \bibfield  {author} {\bibinfo {author} {\bibfnamefont {Dario}\ \bibnamefont
  {Müller}},\ }\bibfield  {title} {\enquote {\bibinfo {title} {{Leptoquarks in
  Flavour Physics}},}\ }in\ \href
  {http://inspirehep.net/record/1647395/files/arXiv:1801.03380.pdf} {\emph
  {\bibinfo {booktitle} {{Workshop on Flavour changing and conserving processes
  (FCCP2017) Anacapri, Capri Island, Italy, September 7-9, 2017}}}}\ (\bibinfo
  {year} {2018})\ \Eprint {http://arxiv.org/abs/1801.03380} {arXiv:1801.03380
  [hep-ph]} \BibitemShut {NoStop}%
\bibitem [{\citenamefont {Doršner}\ and\ \citenamefont
  {Greljo}(2018)}]{Dorsner:2018ynv}%
  \BibitemOpen
  \bibfield  {author} {\bibinfo {author} {\bibfnamefont {Ilja}\ \bibnamefont
  {Doršner}}\ and\ \bibinfo {author} {\bibfnamefont {Admir}\ \bibnamefont
  {Greljo}},\ }\bibfield  {title} {\enquote {\bibinfo {title} {{Leptoquark
  toolbox for precision collider studies}},}\ }\href@noop {} {\  (\bibinfo
  {year} {2018})},\ \Eprint {http://arxiv.org/abs/1801.07641} {arXiv:1801.07641
  [hep-ph]} \BibitemShut {NoStop}%
\bibitem [{\citenamefont {Hiller}\ \emph {et~al.}(2018)\citenamefont {Hiller},
  \citenamefont {Loose},\ and\ \citenamefont {Nišandžić}}]{Hiller:2018wbv}%
  \BibitemOpen
  \bibfield  {author} {\bibinfo {author} {\bibfnamefont {Gudrun}\ \bibnamefont
  {Hiller}}, \bibinfo {author} {\bibfnamefont {Dennis}\ \bibnamefont {Loose}},
  \ and\ \bibinfo {author} {\bibfnamefont {Ivan}\ \bibnamefont
  {Nišandžić}},\ }\bibfield  {title} {\enquote {\bibinfo {title} {{Flavorful
  leptoquarks at hadron colliders}},}\ }\href {\doibase
  10.1103/PhysRevD.97.075004} {\bibfield  {journal} {\bibinfo  {journal} {Phys.
  Rev.}\ }\textbf {\bibinfo {volume} {D97}},\ \bibinfo {pages} {075004}
  (\bibinfo {year} {2018})},\ \Eprint {http://arxiv.org/abs/1801.09399}
  {arXiv:1801.09399 [hep-ph]} \BibitemShut {NoStop}%
\bibitem [{\citenamefont {Fajfer}\ \emph {et~al.}(2018)\citenamefont {Fajfer},
  \citenamefont {Košnik},\ and\ \citenamefont {Vale~Silva}}]{Fajfer:2018bfj}%
  \BibitemOpen
  \bibfield  {author} {\bibinfo {author} {\bibfnamefont {S.}~\bibnamefont
  {Fajfer}}, \bibinfo {author} {\bibfnamefont {N.}~\bibnamefont {Košnik}}, \
  and\ \bibinfo {author} {\bibfnamefont {L.}~\bibnamefont {Vale~Silva}},\
  }\bibfield  {title} {\enquote {\bibinfo {title} {{Footprints of leptoquarks:
  from $ R_{K^{(*)}} $ to $ K \rightarrow \pi \nu \bar{\nu }$}},}\ }\href
  {\doibase 10.1140/epjc/s10052-018-5757-5} {\bibfield  {journal} {\bibinfo
  {journal} {Eur. Phys. J.}\ }\textbf {\bibinfo {volume} {C78}},\ \bibinfo
  {pages} {275} (\bibinfo {year} {2018})},\ \Eprint
  {http://arxiv.org/abs/1802.00786} {arXiv:1802.00786 [hep-ph]} \BibitemShut
  {NoStop}%
\bibitem [{\citenamefont {Monteux}\ and\ \citenamefont
  {Rajaraman}(2018)}]{Monteux:2018ufc}%
  \BibitemOpen
  \bibfield  {author} {\bibinfo {author} {\bibfnamefont {Angelo}\ \bibnamefont
  {Monteux}}\ and\ \bibinfo {author} {\bibfnamefont {Arvind}\ \bibnamefont
  {Rajaraman}},\ }\bibfield  {title} {\enquote {\bibinfo {title} {{B Anomalies
  and Leptoquarks at the LHC: Beyond the Lepton-Quark Final State}},}\
  }\href@noop {} {\  (\bibinfo {year} {2018})},\ \Eprint
  {http://arxiv.org/abs/1803.05962} {arXiv:1803.05962 [hep-ph]} \BibitemShut
  {NoStop}%
\bibitem [{\citenamefont {Arnan}\ \emph {et~al.}(2017)\citenamefont {Arnan},
  \citenamefont {Hofer}, \citenamefont {Mescia},\ and\ \citenamefont
  {Crivellin}}]{Arnan:2016cpy}%
  \BibitemOpen
  \bibfield  {author} {\bibinfo {author} {\bibfnamefont {Pere}\ \bibnamefont
  {Arnan}}, \bibinfo {author} {\bibfnamefont {Lars}\ \bibnamefont {Hofer}},
  \bibinfo {author} {\bibfnamefont {Federico}\ \bibnamefont {Mescia}}, \ and\
  \bibinfo {author} {\bibfnamefont {Andreas}\ \bibnamefont {Crivellin}},\
  }\bibfield  {title} {\enquote {\bibinfo {title} {{Loop effects of heavy new
  scalars and fermions in $b\to s\mu^+\mu^-$}},}\ }\href {\doibase
  10.1007/JHEP04(2017)043} {\bibfield  {journal} {\bibinfo  {journal} {JHEP}\
  }\textbf {\bibinfo {volume} {04}},\ \bibinfo {pages} {043} (\bibinfo {year}
  {2017})},\ \Eprint {http://arxiv.org/abs/1608.07832} {arXiv:1608.07832
  [hep-ph]} \BibitemShut {NoStop}%
\bibitem [{\citenamefont {Gripaios}\ \emph {et~al.}(2016)\citenamefont
  {Gripaios}, \citenamefont {Nardecchia},\ and\ \citenamefont
  {Renner}}]{Gripaios:2015gra}%
  \BibitemOpen
  \bibfield  {author} {\bibinfo {author} {\bibfnamefont {Ben}\ \bibnamefont
  {Gripaios}}, \bibinfo {author} {\bibfnamefont {M.}~\bibnamefont
  {Nardecchia}}, \ and\ \bibinfo {author} {\bibfnamefont {S.~A.}\ \bibnamefont
  {Renner}},\ }\bibfield  {title} {\enquote {\bibinfo {title} {{Linear flavour
  violation and anomalies in B physics}},}\ }\href {\doibase
  10.1007/JHEP06(2016)083} {\bibfield  {journal} {\bibinfo  {journal} {JHEP}\
  }\textbf {\bibinfo {volume} {06}},\ \bibinfo {pages} {083} (\bibinfo {year}
  {2016})},\ \Eprint {http://arxiv.org/abs/1509.05020} {arXiv:1509.05020
  [hep-ph]} \BibitemShut {NoStop}%
\bibitem [{\citenamefont {Aristizabal~Sierra}\ \emph
  {et~al.}(2015)\citenamefont {Aristizabal~Sierra}, \citenamefont {Staub},\
  and\ \citenamefont {Vicente}}]{Sierra:2015fma}%
  \BibitemOpen
  \bibfield  {author} {\bibinfo {author} {\bibfnamefont {D.}~\bibnamefont
  {Aristizabal~Sierra}}, \bibinfo {author} {\bibfnamefont {Florian}\
  \bibnamefont {Staub}}, \ and\ \bibinfo {author} {\bibfnamefont {Avelino}\
  \bibnamefont {Vicente}},\ }\bibfield  {title} {\enquote {\bibinfo {title}
  {{Shedding light on the $b\to s$ anomalies with a dark sector}},}\ }\href
  {\doibase 10.1103/PhysRevD.92.015001} {\bibfield  {journal} {\bibinfo
  {journal} {Phys. Rev.}\ }\textbf {\bibinfo {volume} {D92}},\ \bibinfo {pages}
  {015001} (\bibinfo {year} {2015})},\ \Eprint
  {http://arxiv.org/abs/1503.06077} {arXiv:1503.06077 [hep-ph]} \BibitemShut
  {NoStop}%
\bibitem [{\citenamefont {B\'elanger}\ \emph {et~al.}(2015)\citenamefont
  {B\'elanger}, \citenamefont {Delaunay},\ and\ \citenamefont
  {Westhoff}}]{Belanger:2015nma}%
  \BibitemOpen
  \bibfield  {author} {\bibinfo {author} {\bibfnamefont {Genevi\`eve}\
  \bibnamefont {B\'elanger}}, \bibinfo {author} {\bibfnamefont {C\'edric}\
  \bibnamefont {Delaunay}}, \ and\ \bibinfo {author} {\bibfnamefont {Susanne}\
  \bibnamefont {Westhoff}},\ }\bibfield  {title} {\enquote {\bibinfo {title}
  {{A Dark Matter Relic From Muon Anomalies}},}\ }\href {\doibase
  10.1103/PhysRevD.92.055021} {\bibfield  {journal} {\bibinfo  {journal} {Phys.
  Rev.}\ }\textbf {\bibinfo {volume} {D92}},\ \bibinfo {pages} {055021}
  (\bibinfo {year} {2015})},\ \Eprint {http://arxiv.org/abs/1507.06660}
  {arXiv:1507.06660 [hep-ph]} \BibitemShut {NoStop}%
\bibitem [{\citenamefont {Celis}\ \emph {et~al.}(2017)\citenamefont {Celis},
  \citenamefont {Feng},\ and\ \citenamefont {Vollmann}}]{Celis:2016ayl}%
  \BibitemOpen
  \bibfield  {author} {\bibinfo {author} {\bibfnamefont {Alejandro}\
  \bibnamefont {Celis}}, \bibinfo {author} {\bibfnamefont {Wan-Zhe}\
  \bibnamefont {Feng}}, \ and\ \bibinfo {author} {\bibfnamefont {Martin}\
  \bibnamefont {Vollmann}},\ }\bibfield  {title} {\enquote {\bibinfo {title}
  {{Dirac dark matter and $b \to s \ell^+ \ell^-$ with $\mathrm{U(1)}$ gauge
  symmetry}},}\ }\href {\doibase 10.1103/PhysRevD.95.035018} {\bibfield
  {journal} {\bibinfo  {journal} {Phys. Rev.}\ }\textbf {\bibinfo {volume}
  {D95}},\ \bibinfo {pages} {035018} (\bibinfo {year} {2017})},\ \Eprint
  {http://arxiv.org/abs/1608.03894} {arXiv:1608.03894 [hep-ph]} \BibitemShut
  {NoStop}%
\bibitem [{\citenamefont {Altmannshofer}\ \emph {et~al.}(2016)\citenamefont
  {Altmannshofer}, \citenamefont {Gori}, \citenamefont {Profumo},\ and\
  \citenamefont {Queiroz}}]{Altmannshofer:2016jzy}%
  \BibitemOpen
  \bibfield  {author} {\bibinfo {author} {\bibfnamefont {Wolfgang}\
  \bibnamefont {Altmannshofer}}, \bibinfo {author} {\bibfnamefont {Stefania}\
  \bibnamefont {Gori}}, \bibinfo {author} {\bibfnamefont {Stefano}\
  \bibnamefont {Profumo}}, \ and\ \bibinfo {author} {\bibfnamefont
  {Farinaldo~S.}\ \bibnamefont {Queiroz}},\ }\bibfield  {title} {\enquote
  {\bibinfo {title} {{Explaining dark matter and B decay anomalies with an
  $L_\mu - L_\tau$ model}},}\ }\href {\doibase 10.1007/JHEP12(2016)106}
  {\bibfield  {journal} {\bibinfo  {journal} {JHEP}\ }\textbf {\bibinfo
  {volume} {12}},\ \bibinfo {pages} {106} (\bibinfo {year} {2016})},\ \Eprint
  {http://arxiv.org/abs/1609.04026} {arXiv:1609.04026 [hep-ph]} \BibitemShut
  {NoStop}%
\bibitem [{\citenamefont {Cline}\ \emph {et~al.}(2017)\citenamefont {Cline},
  \citenamefont {Cornell}, \citenamefont {London},\ and\ \citenamefont
  {Watanabe}}]{Cline:2017lvv}%
  \BibitemOpen
  \bibfield  {author} {\bibinfo {author} {\bibfnamefont {James~M.}\
  \bibnamefont {Cline}}, \bibinfo {author} {\bibfnamefont {Jonathan~M.}\
  \bibnamefont {Cornell}}, \bibinfo {author} {\bibfnamefont {David}\
  \bibnamefont {London}}, \ and\ \bibinfo {author} {\bibfnamefont {Ryoutaro}\
  \bibnamefont {Watanabe}},\ }\bibfield  {title} {\enquote {\bibinfo {title}
  {{Hidden sector explanation of $B$-decay and cosmic ray anomalies}},}\ }\href
  {\doibase 10.1103/PhysRevD.95.095015} {\bibfield  {journal} {\bibinfo
  {journal} {Phys. Rev.}\ }\textbf {\bibinfo {volume} {D95}},\ \bibinfo {pages}
  {095015} (\bibinfo {year} {2017})},\ \Eprint
  {http://arxiv.org/abs/1702.00395} {arXiv:1702.00395 [hep-ph]} \BibitemShut
  {NoStop}%
\bibitem [{\citenamefont {Baek}(2017)}]{Baek:2017sew}%
  \BibitemOpen
  \bibfield  {author} {\bibinfo {author} {\bibfnamefont {Seungwon}\
  \bibnamefont {Baek}},\ }\bibfield  {title} {\enquote {\bibinfo {title} {{Dark
  matter contribution to $b\to s \mu^+ \mu^-$ anomaly in local
  $U(1)_{L_\mu-L_\tau}$ model}},}\ }\href@noop {} {\  (\bibinfo {year}
  {2017})},\ \Eprint {http://arxiv.org/abs/1707.04573} {arXiv:1707.04573
  [hep-ph]} \BibitemShut {NoStop}%
\bibitem [{\citenamefont {Cline}(2017)}]{Cline:2017aed}%
  \BibitemOpen
  \bibfield  {author} {\bibinfo {author} {\bibfnamefont {James~M.}\
  \bibnamefont {Cline}},\ }\bibfield  {title} {\enquote {\bibinfo {title} {{$B$
  decay anomalies and dark matter from vectorlike confinement}},}\ }\href@noop
  {} {\  (\bibinfo {year} {2017})},\ \Eprint {http://arxiv.org/abs/1710.02140}
  {arXiv:1710.02140 [hep-ph]} \BibitemShut {NoStop}%
\bibitem [{\citenamefont {Sala}\ and\ \citenamefont
  {Straub}(2017)}]{Sala:2017ihs}%
  \BibitemOpen
  \bibfield  {author} {\bibinfo {author} {\bibfnamefont {Filippo}\ \bibnamefont
  {Sala}}\ and\ \bibinfo {author} {\bibfnamefont {David~M.}\ \bibnamefont
  {Straub}},\ }\bibfield  {title} {\enquote {\bibinfo {title} {{A New Light
  Particle in B Decays?}}}\ }\href {\doibase 10.1016/j.physletb.2017.09.072}
  {\bibfield  {journal} {\bibinfo  {journal} {Phys. Lett.}\ }\textbf {\bibinfo
  {volume} {B774}},\ \bibinfo {pages} {205--209} (\bibinfo {year} {2017})},\
  \Eprint {http://arxiv.org/abs/1704.06188} {arXiv:1704.06188 [hep-ph]}
  \BibitemShut {NoStop}%
\bibitem [{\citenamefont {Kawamura}\ \emph {et~al.}(2017)\citenamefont
  {Kawamura}, \citenamefont {Okawa},\ and\ \citenamefont
  {Omura}}]{Kawamura:2017ecz}%
  \BibitemOpen
  \bibfield  {author} {\bibinfo {author} {\bibfnamefont {Junichiro}\
  \bibnamefont {Kawamura}}, \bibinfo {author} {\bibfnamefont {Shohei}\
  \bibnamefont {Okawa}}, \ and\ \bibinfo {author} {\bibfnamefont {Yuji}\
  \bibnamefont {Omura}},\ }\bibfield  {title} {\enquote {\bibinfo {title}
  {{Interplay between the b$\to s\ell\ell$ anomalies and dark matter
  physics}},}\ }\href {\doibase 10.1103/PhysRevD.96.075041} {\bibfield
  {journal} {\bibinfo  {journal} {Phys. Rev.}\ }\textbf {\bibinfo {volume}
  {D96}},\ \bibinfo {pages} {075041} (\bibinfo {year} {2017})},\ \Eprint
  {http://arxiv.org/abs/1706.04344} {arXiv:1706.04344 [hep-ph]} \BibitemShut
  {NoStop}%
\bibitem [{\citenamefont {Chiang}\ \emph
  {et~al.}(2017{\natexlab{b}})\citenamefont {Chiang}, \citenamefont {Huang},\
  and\ \citenamefont {Okada}}]{Chiang:2017zkh}%
  \BibitemOpen
  \bibfield  {author} {\bibinfo {author} {\bibfnamefont {Cheng-Wei}\
  \bibnamefont {Chiang}}, \bibinfo {author} {\bibfnamefont {Guan-Jie}\
  \bibnamefont {Huang}}, \ and\ \bibinfo {author} {\bibfnamefont {Hiroshi}\
  \bibnamefont {Okada}},\ }\bibfield  {title} {\enquote {\bibinfo {title} {{A
  simple model for explaining muon-related anomalies and dark matter}},}\
  }\href@noop {} {\  (\bibinfo {year} {2017}{\natexlab{b}})},\ \Eprint
  {http://arxiv.org/abs/1711.07365} {arXiv:1711.07365 [hep-ph]} \BibitemShut
  {NoStop}%
\bibitem [{\citenamefont {Bhattacharya}\ \emph {et~al.}(2015)\citenamefont
  {Bhattacharya}, \citenamefont {London}, \citenamefont {Cline}, \citenamefont
  {Datta},\ and\ \citenamefont {Dupuis}}]{Bhattacharya:2015xha}%
  \BibitemOpen
  \bibfield  {author} {\bibinfo {author} {\bibfnamefont {Bhubanjyoti}\
  \bibnamefont {Bhattacharya}}, \bibinfo {author} {\bibfnamefont {David}\
  \bibnamefont {London}}, \bibinfo {author} {\bibfnamefont {James~M.}\
  \bibnamefont {Cline}}, \bibinfo {author} {\bibfnamefont {Alakabha}\
  \bibnamefont {Datta}}, \ and\ \bibinfo {author} {\bibfnamefont {Grace}\
  \bibnamefont {Dupuis}},\ }\bibfield  {title} {\enquote {\bibinfo {title}
  {{Quark-flavored scalar dark matter}},}\ }\href {\doibase
  10.1103/PhysRevD.92.115012} {\bibfield  {journal} {\bibinfo  {journal} {Phys.
  Rev.}\ }\textbf {\bibinfo {volume} {D92}},\ \bibinfo {pages} {115012}
  (\bibinfo {year} {2015})},\ \Eprint {http://arxiv.org/abs/1509.04271}
  {arXiv:1509.04271 [hep-ph]} \BibitemShut {NoStop}%
\bibitem [{\citenamefont {Bona}\ \emph {et~al.}(2008)\citenamefont {Bona} \emph
  {et~al.}}]{Bona:2007vi}%
  \BibitemOpen
  \bibfield  {author} {\bibinfo {author} {\bibfnamefont {M.}~\bibnamefont
  {Bona}} \emph {et~al.} (\bibinfo {collaboration} {UTfit}),\ }\bibfield
  {title} {\enquote {\bibinfo {title} {{Model-independent constraints on
  $\Delta F=2$ operators and the scale of new physics}},}\ }\href {\doibase
  10.1088/1126-6708/2008/03/049} {\bibfield  {journal} {\bibinfo  {journal}
  {JHEP}\ }\textbf {\bibinfo {volume} {03}},\ \bibinfo {pages} {049} (\bibinfo
  {year} {2008})},\ \Eprint {http://arxiv.org/abs/0707.0636} {arXiv:0707.0636
  [hep-ph]} \BibitemShut {NoStop}%
\bibitem [{\citenamefont {Bona}(2016)}]{Bona:2016bvr}%
  \BibitemOpen
  \bibfield  {author} {\bibinfo {author} {\bibfnamefont {Marcella}\
  \bibnamefont {Bona}} (\bibinfo {collaboration} {UTfit}),\ }\bibfield  {title}
  {\enquote {\bibinfo {title} {{Unitarity Triangle analysis beyond the Standard
  Model from UTfit}},}\ }\bibfield  {booktitle} {\emph {\bibinfo {booktitle}
  {{Proceedings, 38th International Conference on High Energy Physics (ICHEP
  2016): Chicago, IL, USA, August 3-10, 2016}}},\ }\href@noop {} {\bibfield
  {journal} {\bibinfo  {journal} {PoS}\ }\textbf {\bibinfo {volume}
  {ICHEP2016}},\ \bibinfo {pages} {149} (\bibinfo {year} {2016})}\BibitemShut
  {NoStop}%
\bibitem [{\citenamefont {Luo}\ and\ \citenamefont {Xiao}(2003)}]{Luo:2002ey}%
  \BibitemOpen
  \bibfield  {author} {\bibinfo {author} {\bibfnamefont {Ming-xing}\
  \bibnamefont {Luo}}\ and\ \bibinfo {author} {\bibfnamefont {Yong}\
  \bibnamefont {Xiao}},\ }\bibfield  {title} {\enquote {\bibinfo {title} {{Two
  loop renormalization group equations in the standard model}},}\ }\href
  {\doibase 10.1103/PhysRevLett.90.011601} {\bibfield  {journal} {\bibinfo
  {journal} {Phys. Rev. Lett.}\ }\textbf {\bibinfo {volume} {90}},\ \bibinfo
  {pages} {011601} (\bibinfo {year} {2003})},\ \Eprint
  {http://arxiv.org/abs/hep-ph/0207271} {arXiv:hep-ph/0207271 [hep-ph]}
  \BibitemShut {NoStop}%
\bibitem [{\citenamefont {Machacek}\ and\ \citenamefont
  {Vaughn}(1983)}]{Machacek:1983tz}%
  \BibitemOpen
  \bibfield  {author} {\bibinfo {author} {\bibfnamefont {Marie~E.}\
  \bibnamefont {Machacek}}\ and\ \bibinfo {author} {\bibfnamefont {Michael~T.}\
  \bibnamefont {Vaughn}},\ }\bibfield  {title} {\enquote {\bibinfo {title}
  {{Two Loop Renormalization Group Equations in a General Quantum Field Theory.
  1. Wave Function Renormalization}},}\ }\href {\doibase
  10.1016/0550-3213(83)90610-7} {\bibfield  {journal} {\bibinfo  {journal}
  {Nucl. Phys.}\ }\textbf {\bibinfo {volume} {B222}},\ \bibinfo {pages}
  {83--103} (\bibinfo {year} {1983})}\BibitemShut {NoStop}%
\bibitem [{\citenamefont {Patrignani}\ \emph {et~al.}(2016)\citenamefont
  {Patrignani} \emph {et~al.}}]{Patrignani:2016xqp}%
  \BibitemOpen
  \bibfield  {author} {\bibinfo {author} {\bibfnamefont {C.}~\bibnamefont
  {Patrignani}} \emph {et~al.} (\bibinfo {collaboration} {Particle Data
  Group}),\ }\bibfield  {title} {\enquote {\bibinfo {title} {{Review of
  Particle Physics}},}\ }\href {\doibase 10.1088/1674-1137/40/10/100001}
  {\bibfield  {journal} {\bibinfo  {journal} {Chin. Phys.}\ }\textbf {\bibinfo
  {volume} {C40}},\ \bibinfo {pages} {100001} (\bibinfo {year}
  {2016})}\BibitemShut {NoStop}%
\bibitem [{\citenamefont {Schael}\ \emph {et~al.}(2006)\citenamefont {Schael}
  \emph {et~al.}}]{ALEPH:2005ab}%
  \BibitemOpen
  \bibfield  {author} {\bibinfo {author} {\bibfnamefont {S.}~\bibnamefont
  {Schael}} \emph {et~al.} (\bibinfo {collaboration} {SLD Electroweak Group,
  DELPHI, ALEPH, SLD, SLD Heavy Flavour Group, OPAL, LEP Electroweak Working
  Group, L3}),\ }\bibfield  {title} {\enquote {\bibinfo {title} {{Precision
  electroweak measurements on the $Z$ resonance}},}\ }\href {\doibase
  10.1016/j.physrep.2005.12.006} {\bibfield  {journal} {\bibinfo  {journal}
  {Phys. Rept.}\ }\textbf {\bibinfo {volume} {427}},\ \bibinfo {pages}
  {257--454} (\bibinfo {year} {2006})},\ \Eprint
  {http://arxiv.org/abs/hep-ex/0509008} {arXiv:hep-ex/0509008 [hep-ex]}
  \BibitemShut {NoStop}%
\bibitem [{\citenamefont {Giunti}\ and\ \citenamefont
  {Studenikin}(2015)}]{Giunti:2014ixa}%
  \BibitemOpen
  \bibfield  {author} {\bibinfo {author} {\bibfnamefont {Carlo}\ \bibnamefont
  {Giunti}}\ and\ \bibinfo {author} {\bibfnamefont {Alexander}\ \bibnamefont
  {Studenikin}},\ }\bibfield  {title} {\enquote {\bibinfo {title} {{Neutrino
  electromagnetic interactions: a window to new physics}},}\ }\href {\doibase
  10.1103/RevModPhys.87.531} {\bibfield  {journal} {\bibinfo  {journal} {Rev.
  Mod. Phys.}\ }\textbf {\bibinfo {volume} {87}},\ \bibinfo {pages} {531}
  (\bibinfo {year} {2015})},\ \Eprint {http://arxiv.org/abs/1403.6344}
  {arXiv:1403.6344 [hep-ph]} \BibitemShut {NoStop}%
\bibitem [{\citenamefont {collaboration}(2017)}]{ATLAS:2017uun}%
  \BibitemOpen
  \bibfield  {author} {\bibinfo {author} {\bibfnamefont {The~ATLAS}\
  \bibnamefont {collaboration}} (\bibinfo {collaboration} {ATLAS}),\ }\bibfield
   {title} {\enquote {\bibinfo {title} {{Search for electroweak production of
  supersymmetric particles in the two and three lepton final state at
  $\boldmath{\sqrt{s}=13\,}$TeV with the ATLAS detector}},}\ }\href@noop {}
  {\bibfield  {journal} {\bibinfo  {journal} {ATLAS-CONF-2017-039}\ } (\bibinfo
  {year} {2017})}\BibitemShut {NoStop}%
\bibitem [{\citenamefont {Kopp}\ \emph {et~al.}(2014)\citenamefont {Kopp},
  \citenamefont {Michaels},\ and\ \citenamefont {Smirnov}}]{Kopp:2014tsa}%
  \BibitemOpen
  \bibfield  {author} {\bibinfo {author} {\bibfnamefont {Joachim}\ \bibnamefont
  {Kopp}}, \bibinfo {author} {\bibfnamefont {Lisa}\ \bibnamefont {Michaels}}, \
  and\ \bibinfo {author} {\bibfnamefont {Juri}\ \bibnamefont {Smirnov}},\
  }\bibfield  {title} {\enquote {\bibinfo {title} {{Loopy Constraints on
  Leptophilic Dark Matter and Internal Bremsstrahlung}},}\ }\href {\doibase
  10.1088/1475-7516/2014/04/022} {\bibfield  {journal} {\bibinfo  {journal}
  {JCAP}\ }\textbf {\bibinfo {volume} {1404}},\ \bibinfo {pages} {022}
  (\bibinfo {year} {2014})},\ \Eprint {http://arxiv.org/abs/1401.6457}
  {arXiv:1401.6457 [hep-ph]} \BibitemShut {NoStop}%
\bibitem [{\citenamefont {Akerib}\ \emph {et~al.}(2017)\citenamefont {Akerib}
  \emph {et~al.}}]{Akerib:2016vxi}%
  \BibitemOpen
  \bibfield  {author} {\bibinfo {author} {\bibfnamefont {D.~S.}\ \bibnamefont
  {Akerib}} \emph {et~al.} (\bibinfo {collaboration} {LUX}),\ }\bibfield
  {title} {\enquote {\bibinfo {title} {{Results from a search for dark matter
  in the complete LUX exposure}},}\ }\href {\doibase
  10.1103/PhysRevLett.118.021303} {\bibfield  {journal} {\bibinfo  {journal}
  {Phys. Rev. Lett.}\ }\textbf {\bibinfo {volume} {118}},\ \bibinfo {pages}
  {021303} (\bibinfo {year} {2017})},\ \Eprint
  {http://arxiv.org/abs/1608.07648} {arXiv:1608.07648 [astro-ph.CO]}
  \BibitemShut {NoStop}%
\bibitem [{\citenamefont {Aalbers}\ \emph {et~al.}(2016)\citenamefont {Aalbers}
  \emph {et~al.}}]{Aalbers:2016jon}%
  \BibitemOpen
  \bibfield  {author} {\bibinfo {author} {\bibfnamefont {J.}~\bibnamefont
  {Aalbers}} \emph {et~al.} (\bibinfo {collaboration} {DARWIN}),\ }\bibfield
  {title} {\enquote {\bibinfo {title} {{DARWIN: towards the ultimate dark
  matter detector}},}\ }\href {\doibase 10.1088/1475-7516/2016/11/017}
  {\bibfield  {journal} {\bibinfo  {journal} {JCAP}\ }\textbf {\bibinfo
  {volume} {1611}},\ \bibinfo {pages} {017} (\bibinfo {year} {2016})},\ \Eprint
  {http://arxiv.org/abs/1606.07001} {arXiv:1606.07001 [astro-ph.IM]}
  \BibitemShut {NoStop}%
\bibitem [{\citenamefont {Chang}\ \emph {et~al.}(2014)\citenamefont {Chang},
  \citenamefont {Edezhath}, \citenamefont {Hutchinson},\ and\ \citenamefont
  {Luty}}]{Chang:2014tea}%
  \BibitemOpen
  \bibfield  {author} {\bibinfo {author} {\bibfnamefont {Spencer}\ \bibnamefont
  {Chang}}, \bibinfo {author} {\bibfnamefont {Ralph}\ \bibnamefont {Edezhath}},
  \bibinfo {author} {\bibfnamefont {Jeffrey}\ \bibnamefont {Hutchinson}}, \
  and\ \bibinfo {author} {\bibfnamefont {Markus}\ \bibnamefont {Luty}},\
  }\bibfield  {title} {\enquote {\bibinfo {title} {{Leptophilic Effective
  WIMPs}},}\ }\href {\doibase 10.1103/PhysRevD.90.015011} {\bibfield  {journal}
  {\bibinfo  {journal} {Phys. Rev.}\ }\textbf {\bibinfo {volume} {D90}},\
  \bibinfo {pages} {015011} (\bibinfo {year} {2014})},\ \Eprint
  {http://arxiv.org/abs/1402.7358} {arXiv:1402.7358 [hep-ph]} \BibitemShut
  {NoStop}%
\bibitem [{\citenamefont {Ade}\ \emph {et~al.}(2016)\citenamefont {Ade} \emph
  {et~al.}}]{Ade:2015xua}%
  \BibitemOpen
  \bibfield  {author} {\bibinfo {author} {\bibfnamefont {P.~A.~R.}\
  \bibnamefont {Ade}} \emph {et~al.} (\bibinfo {collaboration} {Planck}),\
  }\bibfield  {title} {\enquote {\bibinfo {title} {{Planck 2015 results. XIII.
  Cosmological parameters}},}\ }\href {\doibase 10.1051/0004-6361/201525830}
  {\bibfield  {journal} {\bibinfo  {journal} {Astron. Astrophys.}\ }\textbf
  {\bibinfo {volume} {594}},\ \bibinfo {pages} {A13} (\bibinfo {year}
  {2016})},\ \Eprint {http://arxiv.org/abs/1502.01589} {arXiv:1502.01589
  [astro-ph.CO]} \BibitemShut {NoStop}%
\bibitem [{\citenamefont {Steigman}\ \emph {et~al.}(2012)\citenamefont
  {Steigman}, \citenamefont {Dasgupta},\ and\ \citenamefont
  {Beacom}}]{Steigman:2012nb}%
  \BibitemOpen
  \bibfield  {author} {\bibinfo {author} {\bibfnamefont {Gary}\ \bibnamefont
  {Steigman}}, \bibinfo {author} {\bibfnamefont {Basudeb}\ \bibnamefont
  {Dasgupta}}, \ and\ \bibinfo {author} {\bibfnamefont {John~F.}\ \bibnamefont
  {Beacom}},\ }\bibfield  {title} {\enquote {\bibinfo {title} {{Precise Relic
  WIMP Abundance and its Impact on Searches for Dark Matter Annihilation}},}\
  }\href {\doibase 10.1103/PhysRevD.86.023506} {\bibfield  {journal} {\bibinfo
  {journal} {Phys. Rev.}\ }\textbf {\bibinfo {volume} {D86}},\ \bibinfo {pages}
  {023506} (\bibinfo {year} {2012})},\ \Eprint {http://arxiv.org/abs/1204.3622}
  {arXiv:1204.3622 [hep-ph]} \BibitemShut {NoStop}%
\bibitem [{\citenamefont {Belanger}\ \emph {et~al.}(2007)\citenamefont
  {Belanger}, \citenamefont {Boudjema}, \citenamefont {Pukhov},\ and\
  \citenamefont {Semenov}}]{Belanger:2006is}%
  \BibitemOpen
  \bibfield  {author} {\bibinfo {author} {\bibfnamefont {G.}~\bibnamefont
  {Belanger}}, \bibinfo {author} {\bibfnamefont {F.}~\bibnamefont {Boudjema}},
  \bibinfo {author} {\bibfnamefont {A.}~\bibnamefont {Pukhov}}, \ and\ \bibinfo
  {author} {\bibfnamefont {A.}~\bibnamefont {Semenov}},\ }\bibfield  {title}
  {\enquote {\bibinfo {title} {{MicrOMEGAs 2.0: A Program to calculate the
  relic density of dark matter in a generic model}},}\ }\href {\doibase
  10.1016/j.cpc.2006.11.008} {\bibfield  {journal} {\bibinfo  {journal}
  {Comput. Phys. Commun.}\ }\textbf {\bibinfo {volume} {176}},\ \bibinfo
  {pages} {367--382} (\bibinfo {year} {2007})},\ \Eprint
  {http://arxiv.org/abs/hep-ph/0607059} {arXiv:hep-ph/0607059 [hep-ph]}
  \BibitemShut {NoStop}%
\bibitem [{\citenamefont {Belanger}\ \emph {et~al.}(2009)\citenamefont
  {Belanger}, \citenamefont {Boudjema}, \citenamefont {Pukhov},\ and\
  \citenamefont {Semenov}}]{Belanger:2008sj}%
  \BibitemOpen
  \bibfield  {author} {\bibinfo {author} {\bibfnamefont {G.}~\bibnamefont
  {Belanger}}, \bibinfo {author} {\bibfnamefont {F.}~\bibnamefont {Boudjema}},
  \bibinfo {author} {\bibfnamefont {A.}~\bibnamefont {Pukhov}}, \ and\ \bibinfo
  {author} {\bibfnamefont {A.}~\bibnamefont {Semenov}},\ }\bibfield  {title}
  {\enquote {\bibinfo {title} {{Dark matter direct detection rate in a generic
  model with micrOMEGAs 2.2}},}\ }\href {\doibase 10.1016/j.cpc.2008.11.019}
  {\bibfield  {journal} {\bibinfo  {journal} {Comput. Phys. Commun.}\ }\textbf
  {\bibinfo {volume} {180}},\ \bibinfo {pages} {747--767} (\bibinfo {year}
  {2009})},\ \Eprint {http://arxiv.org/abs/0803.2360} {arXiv:0803.2360
  [hep-ph]} \BibitemShut {NoStop}%
\bibitem [{\citenamefont {Bringmann}\ \emph {et~al.}(2017)\citenamefont
  {Bringmann} \emph {et~al.}}]{Workgroup:2017lvb}%
  \BibitemOpen
  \bibfield  {author} {\bibinfo {author} {\bibfnamefont {Torsten}\ \bibnamefont
  {Bringmann}} \emph {et~al.},\ }\bibfield  {title} {\enquote {\bibinfo {title}
  {{DarkBit: A GAMBIT module for computing dark matter observables and
  likelihoods}},}\ }\href {\doibase 10.1140/epjc/s10052-017-5155-4} {\bibfield
  {journal} {\bibinfo  {journal} {Eur. Phys. J.}\ }\textbf {\bibinfo {volume}
  {C77}},\ \bibinfo {pages} {831} (\bibinfo {year} {2017})},\ \Eprint
  {http://arxiv.org/abs/1705.07920} {arXiv:1705.07920 [hep-ph]} \BibitemShut
  {NoStop}%
\bibitem [{\citenamefont {Amole}\ \emph {et~al.}(2017)\citenamefont {Amole}
  \emph {et~al.}}]{Amole:2017dex}%
  \BibitemOpen
  \bibfield  {author} {\bibinfo {author} {\bibfnamefont {C.}~\bibnamefont
  {Amole}} \emph {et~al.} (\bibinfo {collaboration} {PICO}),\ }\bibfield
  {title} {\enquote {\bibinfo {title} {{Dark Matter Search Results from the
  PICO-60 C$_3$F$_8$ Bubble Chamber}},}\ }\href {\doibase
  10.1103/PhysRevLett.118.251301} {\bibfield  {journal} {\bibinfo  {journal}
  {Phys. Rev. Lett.}\ }\textbf {\bibinfo {volume} {118}},\ \bibinfo {pages}
  {251301} (\bibinfo {year} {2017})},\ \Eprint
  {http://arxiv.org/abs/1702.07666} {arXiv:1702.07666 [astro-ph.CO]}
  \BibitemShut {NoStop}%
\bibitem [{\citenamefont {Szydagis}(2016)}]{Szydagis:2016few}%
  \BibitemOpen
  \bibfield  {author} {\bibinfo {author} {\bibfnamefont {M.}~\bibnamefont
  {Szydagis}} (\bibinfo {collaboration} {LUX, LZ}),\ }\bibfield  {title}
  {\enquote {\bibinfo {title} {{The Present and Future of Searching for Dark
  Matter with LUX and LZ}},}\ }\bibfield  {booktitle} {\emph {\bibinfo
  {booktitle} {{Proceedings, 38th International Conference on High Energy
  Physics (ICHEP 2016): Chicago, IL, USA, August 3-10, 2016}}},\ }\href@noop {}
  {\bibfield  {journal} {\bibinfo  {journal} {PoS}\ }\textbf {\bibinfo {volume}
  {ICHEP2016}},\ \bibinfo {pages} {220} (\bibinfo {year} {2016})},\ \Eprint
  {http://arxiv.org/abs/1611.05525} {arXiv:1611.05525 [astro-ph.CO]}
  \BibitemShut {NoStop}%
\bibitem [{\citenamefont {collaboration}(2016{\natexlab{a}})}]{ATLAS:2016ljb}%
  \BibitemOpen
  \bibfield  {author} {\bibinfo {author} {\bibfnamefont {The~ATLAS}\
  \bibnamefont {collaboration}} (\bibinfo {collaboration} {ATLAS}),\ }\bibfield
   {title} {\enquote {\bibinfo {title} {{Search for top squarks in final states
  with one isolated lepton, jets, and missing transverse momentum in $\sqrt{s}$
  = 13 TeV pp collisions with the ATLAS detector}},}\ }\href@noop {} {\bibfield
   {journal} {\bibinfo  {journal} {ATLAS-CONF-2016-050}\ } (\bibinfo {year}
  {2016}{\natexlab{a}})}\BibitemShut {NoStop}%
\bibitem [{\citenamefont {collaboration}(2016{\natexlab{b}})}]{ATLAS:2016xcm}%
  \BibitemOpen
  \bibfield  {author} {\bibinfo {author} {\bibfnamefont {The~ATLAS}\
  \bibnamefont {collaboration}} (\bibinfo {collaboration} {ATLAS}),\ }\bibfield
   {title} {\enquote {\bibinfo {title} {{Search for direct top squark pair
  production and dark matter production in final states with two leptons in
  $\sqrt{s} = 13$ TeV $pp$ collisions using 13.3 fb$^{-1}$ of ATLAS data}},}\
  }\href@noop {} {\bibfield  {journal} {\bibinfo  {journal}
  {ATLAS-CONF-2016-076}\ } (\bibinfo {year} {2016}{\natexlab{b}})}\BibitemShut
  {NoStop}%
\bibitem [{\citenamefont {Alwall}\ \emph {et~al.}(2014)\citenamefont {Alwall},
  \citenamefont {Frederix}, \citenamefont {Frixione}, \citenamefont {Hirschi},
  \citenamefont {Maltoni}, \citenamefont {Mattelaer}, \citenamefont {Shao},
  \citenamefont {Stelzer}, \citenamefont {Torrielli},\ and\ \citenamefont
  {Zaro}}]{Alwall:2014hca}%
  \BibitemOpen
  \bibfield  {author} {\bibinfo {author} {\bibfnamefont {J.}~\bibnamefont
  {Alwall}}, \bibinfo {author} {\bibfnamefont {R.}~\bibnamefont {Frederix}},
  \bibinfo {author} {\bibfnamefont {S.}~\bibnamefont {Frixione}}, \bibinfo
  {author} {\bibfnamefont {V.}~\bibnamefont {Hirschi}}, \bibinfo {author}
  {\bibfnamefont {F.}~\bibnamefont {Maltoni}}, \bibinfo {author} {\bibfnamefont
  {O.}~\bibnamefont {Mattelaer}}, \bibinfo {author} {\bibfnamefont {H.~S.}\
  \bibnamefont {Shao}}, \bibinfo {author} {\bibfnamefont {T.}~\bibnamefont
  {Stelzer}}, \bibinfo {author} {\bibfnamefont {P.}~\bibnamefont {Torrielli}},
  \ and\ \bibinfo {author} {\bibfnamefont {M.}~\bibnamefont {Zaro}},\
  }\bibfield  {title} {\enquote {\bibinfo {title} {{The automated computation
  of tree-level and next-to-leading order differential cross sections, and
  their matching to parton shower simulations}},}\ }\href {\doibase
  10.1007/JHEP07(2014)079} {\bibfield  {journal} {\bibinfo  {journal} {JHEP}\
  }\textbf {\bibinfo {volume} {07}},\ \bibinfo {pages} {079} (\bibinfo {year}
  {2014})},\ \Eprint {http://arxiv.org/abs/1405.0301} {arXiv:1405.0301
  [hep-ph]} \BibitemShut {NoStop}%
\bibitem [{\citenamefont {Drees}\ \emph {et~al.}(2015)\citenamefont {Drees},
  \citenamefont {Dreiner}, \citenamefont {Schmeier}, \citenamefont
  {Tattersall},\ and\ \citenamefont {Kim}}]{Drees:2013wra}%
  \BibitemOpen
  \bibfield  {author} {\bibinfo {author} {\bibfnamefont {Manuel}\ \bibnamefont
  {Drees}}, \bibinfo {author} {\bibfnamefont {Herbi}\ \bibnamefont {Dreiner}},
  \bibinfo {author} {\bibfnamefont {Daniel}\ \bibnamefont {Schmeier}}, \bibinfo
  {author} {\bibfnamefont {Jamie}\ \bibnamefont {Tattersall}}, \ and\ \bibinfo
  {author} {\bibfnamefont {Jong~Soo}\ \bibnamefont {Kim}},\ }\bibfield  {title}
  {\enquote {\bibinfo {title} {{CheckMATE: Confronting your Favourite New
  Physics Model with LHC Data}},}\ }\href {\doibase 10.1016/j.cpc.2014.10.018}
  {\bibfield  {journal} {\bibinfo  {journal} {Comput. Phys. Commun.}\ }\textbf
  {\bibinfo {volume} {187}},\ \bibinfo {pages} {227--265} (\bibinfo {year}
  {2015})},\ \Eprint {http://arxiv.org/abs/1312.2591} {arXiv:1312.2591
  [hep-ph]} \BibitemShut {NoStop}%
\bibitem [{\citenamefont {Alloul}\ \emph {et~al.}(2014)\citenamefont {Alloul},
  \citenamefont {Christensen}, \citenamefont {Degrande}, \citenamefont {Duhr},\
  and\ \citenamefont {Fuks}}]{Alloul:2013bka}%
  \BibitemOpen
  \bibfield  {author} {\bibinfo {author} {\bibfnamefont {Adam}\ \bibnamefont
  {Alloul}}, \bibinfo {author} {\bibfnamefont {Neil~D.}\ \bibnamefont
  {Christensen}}, \bibinfo {author} {\bibfnamefont {C\'eline}\ \bibnamefont
  {Degrande}}, \bibinfo {author} {\bibfnamefont {Claude}\ \bibnamefont {Duhr}},
  \ and\ \bibinfo {author} {\bibfnamefont {Benjamin}\ \bibnamefont {Fuks}},\
  }\bibfield  {title} {\enquote {\bibinfo {title} {{FeynRules 2.0 - A complete
  toolbox for tree-level phenomenology}},}\ }\href {\doibase
  10.1016/j.cpc.2014.04.012} {\bibfield  {journal} {\bibinfo  {journal}
  {Comput. Phys. Commun.}\ }\textbf {\bibinfo {volume} {185}},\ \bibinfo
  {pages} {2250--2300} (\bibinfo {year} {2014})},\ \Eprint
  {http://arxiv.org/abs/1310.1921} {arXiv:1310.1921 [hep-ph]} \BibitemShut
  {NoStop}%
\bibitem [{\citenamefont {Sjöstrand}\ \emph {et~al.}(2015)\citenamefont
  {Sjöstrand}, \citenamefont {Ask}, \citenamefont {Christiansen},
  \citenamefont {Corke}, \citenamefont {Desai}, \citenamefont {Ilten},
  \citenamefont {Mrenna}, \citenamefont {Prestel}, \citenamefont {Rasmussen},\
  and\ \citenamefont {Skands}}]{Sjostrand:2014zea}%
  \BibitemOpen
  \bibfield  {author} {\bibinfo {author} {\bibfnamefont {Torbjörn}\
  \bibnamefont {Sjöstrand}}, \bibinfo {author} {\bibfnamefont {Stefan}\
  \bibnamefont {Ask}}, \bibinfo {author} {\bibfnamefont {Jesper~R.}\
  \bibnamefont {Christiansen}}, \bibinfo {author} {\bibfnamefont {Richard}\
  \bibnamefont {Corke}}, \bibinfo {author} {\bibfnamefont {Nishita}\
  \bibnamefont {Desai}}, \bibinfo {author} {\bibfnamefont {Philip}\
  \bibnamefont {Ilten}}, \bibinfo {author} {\bibfnamefont {Stephen}\
  \bibnamefont {Mrenna}}, \bibinfo {author} {\bibfnamefont {Stefan}\
  \bibnamefont {Prestel}}, \bibinfo {author} {\bibfnamefont {Christine~O.}\
  \bibnamefont {Rasmussen}}, \ and\ \bibinfo {author} {\bibfnamefont
  {Peter~Z.}\ \bibnamefont {Skands}},\ }\bibfield  {title} {\enquote {\bibinfo
  {title} {{An Introduction to PYTHIA 8.2}},}\ }\href {\doibase
  10.1016/j.cpc.2015.01.024} {\bibfield  {journal} {\bibinfo  {journal}
  {Comput. Phys. Commun.}\ }\textbf {\bibinfo {volume} {191}},\ \bibinfo
  {pages} {159--177} (\bibinfo {year} {2015})},\ \Eprint
  {http://arxiv.org/abs/1410.3012} {arXiv:1410.3012 [hep-ph]} \BibitemShut
  {NoStop}%
\bibitem [{\citenamefont {de~Favereau}\ \emph {et~al.}(2014)\citenamefont
  {de~Favereau}, \citenamefont {Delaere}, \citenamefont {Demin}, \citenamefont
  {Giammanco}, \citenamefont {Lema\^{i}tre}, \citenamefont {Mertens},\ and\
  \citenamefont {Selvaggi}}]{deFavereau:2013fsa}%
  \BibitemOpen
  \bibfield  {author} {\bibinfo {author} {\bibfnamefont {J.}~\bibnamefont
  {de~Favereau}}, \bibinfo {author} {\bibfnamefont {C.}~\bibnamefont
  {Delaere}}, \bibinfo {author} {\bibfnamefont {P.}~\bibnamefont {Demin}},
  \bibinfo {author} {\bibfnamefont {A.}~\bibnamefont {Giammanco}}, \bibinfo
  {author} {\bibfnamefont {V.}~\bibnamefont {Lema\^{i}tre}}, \bibinfo {author}
  {\bibfnamefont {A.}~\bibnamefont {Mertens}}, \ and\ \bibinfo {author}
  {\bibfnamefont {M.}~\bibnamefont {Selvaggi}} (\bibinfo {collaboration}
  {DELPHES 3}),\ }\bibfield  {title} {\enquote {\bibinfo {title} {{DELPHES 3, A
  modular framework for fast simulation of a generic collider experiment}},}\
  }\href {\doibase 10.1007/JHEP02(2014)057} {\bibfield  {journal} {\bibinfo
  {journal} {JHEP}\ }\textbf {\bibinfo {volume} {02}},\ \bibinfo {pages} {057}
  (\bibinfo {year} {2014})},\ \Eprint {http://arxiv.org/abs/1307.6346}
  {arXiv:1307.6346 [hep-ex]} \BibitemShut {NoStop}%
\bibitem [{\citenamefont {Albrecht}\ \emph {et~al.}(2017)\citenamefont
  {Albrecht}, \citenamefont {Bernlochner}, \citenamefont {Kenzie},
  \citenamefont {Reichert}, \citenamefont {Straub},\ and\ \citenamefont
  {Tully}}]{Albrecht:2017odf}%
  \BibitemOpen
  \bibfield  {author} {\bibinfo {author} {\bibfnamefont {Johannes}\
  \bibnamefont {Albrecht}}, \bibinfo {author} {\bibfnamefont {Florian}\
  \bibnamefont {Bernlochner}}, \bibinfo {author} {\bibfnamefont {Matthew}\
  \bibnamefont {Kenzie}}, \bibinfo {author} {\bibfnamefont {Stefanie}\
  \bibnamefont {Reichert}}, \bibinfo {author} {\bibfnamefont {David}\
  \bibnamefont {Straub}}, \ and\ \bibinfo {author} {\bibfnamefont {Alison}\
  \bibnamefont {Tully}},\ }\bibfield  {title} {\enquote {\bibinfo {title}
  {{Future prospects for exploring present day anomalies in flavour physics
  measurements with Belle II and LHCb}},}\ }\href@noop {} {\  (\bibinfo {year}
  {2017})},\ \Eprint {http://arxiv.org/abs/1709.10308} {arXiv:1709.10308
  [hep-ph]} \BibitemShut {NoStop}%
\end{thebibliography}%

\end{document}